\documentclass{aa}  
\usepackage{graphicx}
\usepackage{txfonts}
\begin{document}

   \title{Properties of planets in binary systems}
   \subtitle{The role of  binary separation}

   \author{S. Desidera
          \inst{1},
           M. Barbieri
          \inst{1,2}}

   \authorrunning{S. Desidera \& M. Barbieri}

   \offprints{S. Desidera,  \\
              \email{silvano.desidera@oapd.inaf.it} }

   \institute{INAF -- Osservatorio Astronomico di Padova,  
              Vicolo dell' Osservatorio 5, I-35122, Padova, Italy
              \and
              Dipartimento di Fisica, Universit\`a di Padova, Italy}

 \date{Received  / Accepted }

\abstract
{}
{The statistical properties of planets in binaries were investigated. Any
 difference 
 to planets orbiting single stars can shed light on the formation and
 evolution of planetary systems.
 As planets were found around components of binaries with very different
 separation and mass ratio, it is particularly important to study the
 characteristics of planets as a function of the effective gravitational
 influence of the companion. }
{A compilation of planets in binary systems was made; a search for
 companions orbiting stars recently shown to host planets was performed,
 resulting in the addition of two further binary planet hosts (HD 20782 and
 HD 109749). The probable original properties of the three binary planet 
 hosts with
 white dwarfs companions were also investigated. Using this updated sample
 of planets in binaries we performed a statistical analysis of
 the distributions of planet mass, period, and eccentricity, fraction
 of multiplanet systems, and stellar metallicity for planets orbiting 
 components of  tight and
 wide binaries and single stars. }
{The only highly significant difference revealed by our analysis
 concerns the mass distribution of short-period planets.
 Massive planets in short period orbits are found in most cases around
 the components of rather tight binaries.
 The properties of exoplanets orbiting the components of wide
 binaries are compatible with those of planets orbiting single stars,
 except for a possible greater abundance of high-eccentricity planets.
 The previously suggested lack of massive planets with $P>100$ days 
 in binaries 
 is not confirmed.}
{We conclude that the presence of a stellar companion with separation
 smaller than 100-300 AU is able to modify the formation and/or 
 migration and/or the dynamical
 evolution history of giant planets while wide companions play 
 a more limited role.}


   \maketitle
%

\section{Introduction}
\label{s:intro}

The study of the frequency of planets in binary systems\footnote{Some of 
the systems considered
in this study are actually hierarchical triple systems. For simplicity, 
we refer to the whole sample as binaries.} and of the 
properties of  these planets and of the binary systems themselves 
is very important to improve our knowledge of planet formation and evolution.
On one hand, the frequency of planets
in binary systems has a strong effect on the global frequency of planets,
more than half of solar type stars being in binary or multiple systems 
(Duquennoy \& Mayor \cite{duq91}).
On the other hand, the properties of planets in binaries, and any
difference to those of the planets orbiting single stars, would
shed light on the effects caused by the presence of the companions.

The search for planets in binaries can follow two 
complementary approaches.
The first one is to perform dedicated surveys looking for planets in 
binary systems.
Several programs are currently in progress focusing on different types of
binaries. The SARG planet search has been studying about 50
wide pairs with similar components (Desidera et al.~\cite{desidera06})
since late 2000.
A similar study was recently started by the Geneva group using HARPS.
About 100 spectroscopic binaries with orbital periods between
1.5 and 100 yr were monitored by 
Eggenberger et al.~(\cite{egg06}), while
Konacki (\cite{konacki05a}) is monitoring about 450 stars 
in binaries.

The second approach is to study the binarity of the hosts of planets discovered
in general surveys, which include many binary stars in their list in spite
of some selection biases against them. Dedicated adaptive optics 
observations are required to reveal faint companions close to the star 
(e.g. Patience et al.~\cite{patience02}; Chauvin et al.~\cite{chauvin06};
Mugrauer et al.~\cite{mugrauer06}; see Table~\ref{t:tablebin} for 
further references). 
However, the search for brighter companions at large separation 
can be performed by checking astronomical catalogs 
for nearby, comoving objects.

The first  analysis of the properties of planets in binaries
revealed differences with respect to those of planets  orbiting single stars.
Zucker \& Mazeh (\cite{zucker02}) showed some difference 
in the period-mass relation.
A more complete statistical analysis by Eggenberger et al.~(\cite{egg04}),
hereafter E04, confirms that massive planets in close orbits are mostly
found in binaries. They also propose other possible peculiar
characteristics of planets in binaries: low eccentricities
for orbital periods shorter than 40 days and a lack of
massive planets with periods longer than 100 days in binaries.
The estimated significance of these latter features is less than  97\%,
therefore requiring  larger samples for confirmation.

The continuing observational efforts of planet searches and
dedicated follow-up  to study the characteristics of planet hosts
lead to a significant increase of the total number of planets
in binaries.
We now know more than 40 planets in binaries or multiple systems,
an increase by more than a factor of two with respect to the 
first compilation of the planets in binary systems  assembled by 
E04 (19 planets in 15 systems\footnote{The radial velocity variations 
of one of the stars listed in E04 (\object{HD 219542B}) was shown
to be due to the activity jitter of the star and not to a planet
(Desidera et al.~\cite{letter_hd219542}).}).

The recent publication of the Catalog of Nearby Exoplanets by Butler
et al.~(\cite{butler06}), hereafter B06, provides an updated and 
homogeneous assembly of planet and stellar properties (excluding binarity).
This work  is a useful starting point for a
statistical
comparison of the properties of planets orbiting the
components of multiple systems and single stars. 
In this paper we present such a discussion.

The outline of the paper is the following.
In Sect.~\ref{s:tablebin}  the samples of planets in binaries and
in single stars are defined; 
in Sect.~\ref{s:analysis} the statistical analysis of the
properties of planets in binaries and single stars is presented.
In Sect.~\ref{s:discussion}  the results and their implications
are discussed.
The appendices \ref{s:newbin}-\ref{s:individuals} address specific 
issues relevant for individual cases:
in Appendix \ref{s:newbin} we present the results of the search 
for companions of stars recently shown to host planets, resulting
in two further binaries (HD 20782 and HD 109749), whose properties
are discussed is detail; in Appendix \ref{s:wd} we discuss the 
probable original
properties (mass ratio, separation) of the three systems including 
white dwarf (hereafter WD) companions; in Appendix \ref{s:individuals}
we provide details
about individual objects.

\section{An updated compilation of planets in binaries}
\label{s:tablebin}

We assembled the properties of the planets in multiple systems
and of the multiple system themselves (Table \ref{t:tablebin}).
For the planet hosts not included in the latest 
study of multiplicity of planet hosts 
(Raghavan et al.~\cite{rag06}, hereafter R06), 
we performed a search in existing catalogs of binaries.
\footnote{Some planets have been announced after the submission
of this paper and they are not considered in this study. 
One of them is orbiting a component of a wide binary
(\object{ADS 16402B}; Bakos et al.~\cite{bakos06b}).
The binary separation is 1550 AU. The two transiting planets
discovered by Collier Cameron et al.~(\cite{wasp}) have companions
candidates  discovered using adaptive-optics
imaging, whose physical association to the planet hosts needs
confirmation.}. 
Details are presented in Appendix \ref{s:newbin}.

The planetary properties, as well as mass and metallicity of the planet hosts,
were taken from B06.
The binary parameters were taken from the references  listed in the 
Table caption.
As we are interested in evaluating the actual dynamical effects of the 
companions,
we listed individual masses, derived using the mass-luminosity relations by 
Delfosse et al.~(\cite{delfosse00}) or taken from the works cited in the 
caption.
In a few cases, the complete 
binary orbit is available, making feasible a more in depth analysis
of the possible formation and evolution of the planet
(see e.g. Thebault et al.~\cite{thebault04} for $\gamma$ Cep).
The last column lists the critical semiaxis for dynamical stability 
of planetary orbit derived
using Eq.~1 of Holman \& Weigert (\cite{holman99}).
When the full  binary orbit is not available,
this was calculated deriving the semimajor axis from the projected
separation and adopting an eccentricity of 0.35 
(Fischer \& Marcy \cite{fischer92}; Duquennoy \& Mayor \cite{duq91}).
Remarks on individual
objects are reported in Appendix \ref{s:individuals}, while the properties
of the systems containing a white dwarf are discussed in 
Appendix \ref{s:wd}.

The comparison sample of planets in single stars is that of B06, excluding the
objects in Table \ref{t:tablebin}.
The mass limit adopted by B06
($m \sin i < 24~M_{J}$) is different  to those
adopted in R06.  
One  'super-planet' in the range $m \sin i~~13-24~M_{J}$ was found in a binary,
the companion orbiting \object{HD 41004B}.

\begin{table*}
   \caption[]{Properties of planets in multiple systems, of their host 
stars and their companions:  planet $ m \sin i $, period, semimajor axis, 
eccentricity, radial velocity semi-amplitude,
 metallicity and mass of the planet host (from B06); projected separation, 
semimajor axis
and eccentricity (when available) and mass of the companions, 
and critical semiaxis for 
dynamical stability of planets. The asterisks in the last column mark systems 
discussed individually in Appendix \ref{s:individuals}. }
     \label{t:tablebin}
      
       \begin{tabular}{llrrrrrrrrrrrrl}
         \hline
         \noalign{\smallskip}
Object  & Object & $ m \sin i $ & P  & $a_{p}$  & $e_{p}$ & $K_{p}$  & [Fe/H] & $M_{star}$           & $\rho$ &  $a_{bin}$ & $e_{bin}$ & $M_{comp}$ & $a_{crit}$  & Rem. \\
  HD    & Other &  $M_J$        & d  & AU            &              &   m/s         &        & $M_{\odot}$ &  AU    &     AU     &           & $M_{\odot}$ & AU & \\

         \noalign{\smallskip}
         \hline
         \noalign{\smallskip}

         142 &                 & 1.31   &  350.3     &  1.045   & 0.26    &    33.9   &  0.100 &  1.24 &     138 &       &        &  0.56   &    35  &  \\
        1237 & GJ 3021         & 3.37   &  133.71    &  0.495   & 0.511   &   167.0   &  0.120 &  0.90 &      70 &       &        &  0.13   &    21  & \\   
        3651 &  54 Psc         & 0.227  &  62.206    &  0.296   & 0.618   &    16.0   &  0.164 &  0.89 &     480 &       &        &  0.06   &   151  & * \\   

        9826 & $\upsilon$ And  & 0.687  &    4.617   &  0.0595  & 0.023   &    69.8   &  0.120 &  1.32 &     750 &       &        &  0.19   &   223  & \\  
             &                 & 1.98   &  241.23    &  0.832   & 0.262   &    55.6   &  0.120 &  1.32 &     750 &       &        &  0.19   &   223  & \\ 
       &                       & 3.95   & 1290.1     &  2.54    & 0.258   &    63.4   &  0.153 &  1.32 &     750 &       &        &  0.19   &   223  & \\  
       11964 &                 & 0.61   & 2110.0     &  3.34    & 0.06    &     9.0   &  0.122 &  1.12 &    1379 &       &        &  0.67   &   325  &  \\  
       13445 & GL 86           & 3.91   &   15.7649  &  0.113   & 0.0416  &   376.7   & -0.268 &  0.77 &         &  18.4 &  0.40  & 0.49&   3.1  & $^a$ \\  
       16141 & 79 Cet          & 0.26   &   75.523   &  0.363   & 0.252   &    11.99  &  0.170 &  1.12 &     223 &       &        &  0.29   &    62  & \\  
       19994 & 94 Cet          & 1.69   &  535.7     &  1.428   & 0.30    &    36.2   &  0.186 &  1.35 &         &  120  &  0.26  &  0.35   &    31  & * \\ 
       20782 &                 & 1.78   &  585.86    &  1.364   & 0.925   &   115.0   & -0.051 &  0.98 &    9080 &       &        &  0.84   &  1940  & \\ 
       27442 & $\epsilon$ Ret  & 1.56   &  428.1     &  1.27    & 0.06    &    32.2   &  0.420 &  1.49 &     251 &       &        & 0.60&    62  & $^a$ \\ 
       38529 &                 & 0.852  &   14.3093  &  0.1313  & 0.248   &    56.8   &  0.445 &  1.47 &   12000 &       &        &  0.50   &  3190  & * \\ 
             &                 &13.2    & 2165.0     &  3.74    & 0.3506  &   170.3   &  0.445 &  1.47 &   12000 &       &        &  0.50   &  3190  & \\
       40979 &                 & 3.83   &  263.84    &  0.855   & 0.269   &   112.0   &  0.168 &  1.19 &    6400 &       &        &  0.75   &  1488  & \\ 
       41004A &                 & 2.60   &  963.0     &  1.70    & 0.74    &    99.0   &  0.160 &  0.70 &     23 &      &         &  0.40   &   5.5  &*  \\   
       41004B &                 &18.4    &    1.3283  &  0.0177  & 0.081   &  6114.0   &  0.160 &  0.40 &     23 &      &         &  0.70   &   3.9  &* \\  
       46375 &                 & 0.226  &    3.0235  &  0.0398  & 0.063   &    33.65  &  0.240 &  0.92 &     346 &       &        &  0.60   &    80  & \\  
       75289 &                 & 0.467  &    3.5092  &  0.0482  & 0.034   &    54.9   &  0.217 &  1.21 &     621 &       &        &  0.14   &   188  & \\ 
       75732 & 55 Cnc          & 0.0377 &    2.7955  &  0.0377  & 0.09    &     5.8   &  0.315 &  0.91 &    1062 &       &        &  0.26   &   291  & \\   
             &                 & 0.833  &   14.652   &  0.1138  & 0.01    &    73.38  &  0.315 &  0.91 &    1062 &       &        &  0.26   &   291  & \\  
             &                 & 0.157  &   44.36    &  0.238   & 0.071   &     9.6   &  0.315 &  0.91 &    1062 &       &        &  0.26   &   291  & \\   
             &                 & 3.90   & 5552.0     &  5.97    & 0.091   &    47.5   &  0.315 &  0.91 &    1062 &       &        &  0.26   &   291  & \\   
       80606 &                 & 4.31   &   111.449  &  0.468   & 0.935   &   481.9   &  0.343 &  1.10 &    1200 &       &        &  0.90   &   260  & \\   
       89744 &                 & 8.58   &  256.80    &  0.934   & 0.677   &   267.3   &  0.265 &  1.64 &    2456 &       &        &  0.079  &   780  &  \\ 
       99492 & 83 Leo B        & 0.109  &   17.0431  &  0.1232  & 0.254   &     9.8   &  0.362 &  0.86 &     515 &       &        &  1.01   &   100  &  \\ 
      109749 &                 & 0.277  &    5.23947 &  0.0629  & 0       &    28.58  &  0.250 &  1.21 &     490 &       &        &  0.78   &   113  & \\ 
      114729 &                 & 0.95   & 1114.0     &  2.11    & 0.167   &    18.8   & -0.262 &  1.00 &     282 &       &        &  0.25   &    79  & \\ 
      114762 &                 &11.68   &   83.8881  &  0.363   & 0.336   &   615.2   & -0.653 &  0.89 &     130 &       &        &  0.07   &    40  & \\ 
      120136 & $\tau$ Boo      & 4.13   &    3.3125  &  0.048   & 0.023   &   461.1   &  0.234 &  1.35 &         &  245  & 0.91   &  0.40   &   2.8  & * \\  
      142022 &                 & 4.50   & 1928.0     &  2.93    & 0.53    &    92.0   &  0.190 &  0.90 &     820 &       &        &  0.60   &   188  & \\ 
      147513 &                 & 1.18   &  528.4     &  1.31    & 0.26    &    29.3   &  0.089 &  1.07 &    4451 &       &        &0.65 &  1044  & $^a$\\
      178911 &                 & 7.35   &   71.511   &  0.345   & 0.139   &   346.9   &  0.285 &  1.06 &     640 &       &        &1.89&   108  & $^b$ \\ 
      186427 & 16 CygB        & 1.68   &  798.5     &  1.681   & 0.681   &    50.5   &  0.038 &  0.99 &     850 &       &        &1.19&   164  & $^b$ \\ 
      188015 &                 & 1.50   &  461.2     &  1.203   & 0.137   &    37.6   &  0.289 &  1.09 &     684 &       &        &  0.21   &   198  &  \\ 
      189733 &                 & 1.15   &    2.219   &  0.0312  & 0       &   205.0   & -0.030 &  0.82 &     216 &       &        &  0.19   &    61  & * \\ 
      190360 &  GJ 777 A       & 0.0587 &   17.1     &  0.1303  & 0.01    &     4.6   &  0.213 &  1.01 &    3000 &       &        &  0.20   &   864  &  \\ 
             &                 & 1.55   & 2891.0     &  3.99    & 0.36    &    23.5   &  0.213 &  1.01 &    3000 &       &        &  0.20   &   864  &  \\
      195019 &                 & 3.69   &   18.2013  &  0.1388  & 0.0138  &   271.5   &  0.068 &  1.07 &     150 &       &        &  0.70   &    35  &  \\ 
      196050 &                 & 2.90   & 1378.0     &  2.454   & 0.228   &    49.7   &  0.229 &  1.15 &     511 &       &        &  0.36   &   138  &  \\  
      213240 &                 & 4.72   &  882.7     &  1.92    & 0.421   &    96.6   &  0.139 &  1.22 &    3898 &       &        &  0.15   &  1177  &  \\ 
      222404 & $\gamma$ Cep    & 1.77   &  905.0     &  2.14    & 0.12    &    27.5   &  0.180 &  1.59 &         &  18.5 &  0.36  &  0.40   &   4.0  &  \\  
      222582 &                 & 7.75   &  572.38    &  1.347   & 0.725   &   276.3   & -0.029 &  0.99 &    4746 &       &        &  0.36   &  1246  &  \\ 

         \noalign{\smallskip}
         \hline
      \end{tabular}

Binary references:
\object{HD 142}: R06;
\object{HD 1237}: Chauvin et al.~(\cite{chauvin06});
\object{HD 3651}: Mugrauer et al.~(\cite{mugrauer06b});
\object{HD 9826}: Lowrance et al.~(\cite{lowrance02}), E04; 
\object{HD 11964}: R06;  
\object{HD 13445}: Lagrange et al.~(\cite{lagrange06}), 
                   Mugrauer \& Neuhauser (\cite{mugneu05});  
\object{HD 16141}: Mugrauer et al.~(\cite{mugrauer05});
\object{HD 19994}: Hale (\cite{hale94}), E04; 
\object{HD 20782}: this paper;
\object{HD 27442}: Chauvin et al.~(\cite{chauvin06}), R06;
\object{HD 38529}: R06;
\object{HD 40979}: E04;
\object{HD 41404}: E04; 
\object{HD 46375}: Mugrauer et al.~(\cite{mugrauer06});
\object{HD 75289}: Mugrauer et al.~(\cite{mugrauer04b}); 
\object{HD 75732}: E04, Mugrauer et al.~(\cite{mugrauer06});   
\object{HD 80606}: E04;  
\object{HD 89744}: Mugrauer et al.~(\cite{mugrauer04a}); 
\object{HD 99492}: R06;
\object{HD 109749}: this paper;  
\object{HD 114729}: Mugrauer et al.~(\cite{mugrauer05});
\object{HD 114762}: Patience et al.~(\cite{patience02}), E04; 
\object{HD 120136}: Hale (\cite{hale94}), E04;
\object{HD 142022}: Eggenberger et al.~(\cite{egg06b});
\object{HD 147513}: Mayor et al.~(\cite{mayor04}), R06, Silvestri et al.~(\cite{silvestri01});
\object{HD 178911}: Tokovinin et al.~(\cite{tok00}), E04; 
\object{HD 186427}: Hauser \& Marcy (\cite{hauser99}), 
                    Patience et al.~(\cite{patience02}), E04; 
\object{HD 188015}: R06;
\object{HD 189733}: Bakos et al.~(\cite{bakos06}); 
\object{HD 190360}: E04;
\object{HD 195019}: E04; 
\object{HD 196050}: Mugrauer et al.~(\cite{mugrauer05});
\object{HD 213240}: Mugrauer et al.~(\cite{mugrauer05});
\object{HD 222404}: Hatzes et al.~(\cite{hatzes03});
\object{HD 222582}: R06.

\begin{list}{}{}
\item[$^{\mathrm{a}}$] The secondary is a white dwarf; the original mass was 
larger and 
the separation smaller. See App.~\ref{s:wd} for details. 
\item[$^{\mathrm{b}}$] Hierarchical triple system, the sum
                       of the masses of the two companions is listed here.
                       See App.~\ref{s:individuals} for details. 
\end{list}

\end{table*}

\section{Statistical analysis}
\label{s:analysis}

\begin{figure}
\includegraphics[width=9cm]{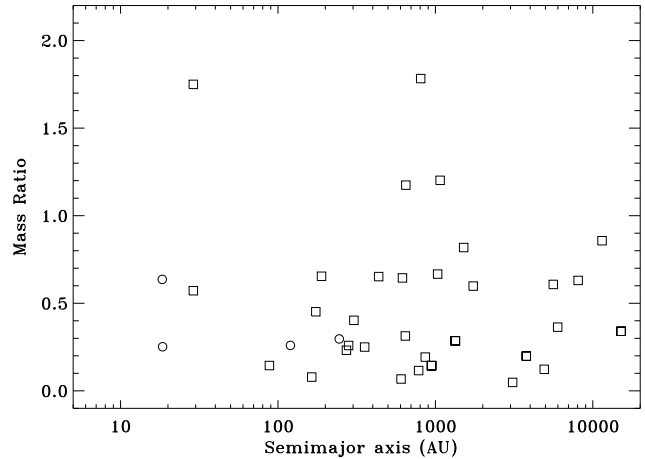}
      \caption{Mass ratio vs semimajor axis of the binary orbit  for 
               stars with planets in binary systems. 
               Open circles represent
               the pairs for which binary orbit is available, open squares
               the pairs for which only the binary separation is available.}        
         \label{f:bin}
   \end{figure}

Fig.~\ref{f:bin} shows the mass ratio vs semimajor axis  for 
stars with planets in multiple systems. For hierarchical triple
systems in which the planet orbits the isolated companion, the masses 
of the binary companions to the planet host 
are summed.
It appears that planets might exist in binaries with very different
properties. In some cases (e.g. very low mass companions at
a projected separation larger than 1000 AU) the dynamical effects
of the companion on the formation and evolution of the planetary system
might be very limited, while in the cases of very tight binaries
the presence  of the planet represents a challenge for the
current models of planet formation (Hatzes \& Wuchterl \cite{hatzes05}).

A simple look at Fig.~\ref{f:bin} suggests a possible depopulated region
for the binary semimajor axis between about 20 to 100 AU.
A proper evaluation of its reality requires a detailed study of the selection
effects of the original samples of radial velocity surveys
and the discovery of stellar companions of planet hosts, which is 
postponed for a future study.

The dynamical effects of the companions on the circumstellar region
of the planet hosts are certainly very different for the binaries
in our sample.
To consider the effects of dynamical perturbation by the
stellar companion(s) we used the critical semiaxis for dynamical
stability of the planet $a_{crit}$ (Holman \& Wiegert \cite{holman99}). 
This allows us to take into account the effects
of both the separation and the mass ratio.
The critical semiaxis for dynamical stability represents the limit
for stable planetary orbits around the planet hosts (calculated
for coplanar circular orbits).
It is also similar to the radius of tidal truncation of the circumstellar disk
(Pichardo et al.~\cite{pichardo05}; Pfahl \& Muterspaugh \cite{pfahl06}).
However, the area in which relative
impact velocities between planetesimals is expected to allow planet formation
might be significantly smaller than $a_{crit}$ (Thebault et al.~\cite{thebault06}).

The critical semimajor axis $a_{crit}$ was used to divide the sample according
to the relevance of the dynamical effects. 
We define as 'tight' binaries those with $a_{crit}<75$~AU
and 'wide' binaries those with $a_{crit}>75$~AU. The limit corresponds to
a projected separation of about 200-300 AU depending on the mass ratio.
The choice  was fixed as a compromise, to obtain 
a sample of binaries wide enough to guarantee limited
dynamical perturbations for the portion of the disk
on which the formation of giant planets
is expected to occur 
and to allow  the inclusion of an adequate number of objects in the
'tight' binary sample.
When relevant for the statistical analysis, the effect of changing
this limit is considered in the following discussion.

The statistical comparison to test the hypothesis that the
parameters of planets (mass, period, eccentricity) in tight and wide binaries 
and in single stars
can be drawn from the same parent distribution was estimated 
using the Kolmogorov-Smirnov (hereafter KS) test and the Mann-Whitney U 
(hereafter MWU) test (Babu \& Feigelson \cite{babu}).
We also used the  hypergeometric
distribution (see E04), that yields the probability of obtaining by chance 
a number of cases from a subsample of given size.
The statistical analysis was performed including only planets
with radial velocity (hereafter RV) semi-amplitude $K>15$ m/s,
to exclude planets with uncertain orbital parameters and to ensure
a more homogeneous detectability. This choice has little effect on
the results.

\subsection{The mass distribution of close-in planets}
\label{s:51peg}

Zucker \& Mazeh (\cite{zucker02}) showed that planets in binaries 
follow a different
mass period relation than those orbiting single stars. 
In particular, high-mass, 
short-period planets appear to be present only in binary systems.
Such a correlation was confirmed by E04 but further inclusion of new systems
makes it weaker (Mugrauer et al.~\cite{mugrauer05}).
The addition of further low-mass hot-Jupiters in binaries from
our work  consolidates this latter trend.

\begin{figure}
   \includegraphics[width=9cm]{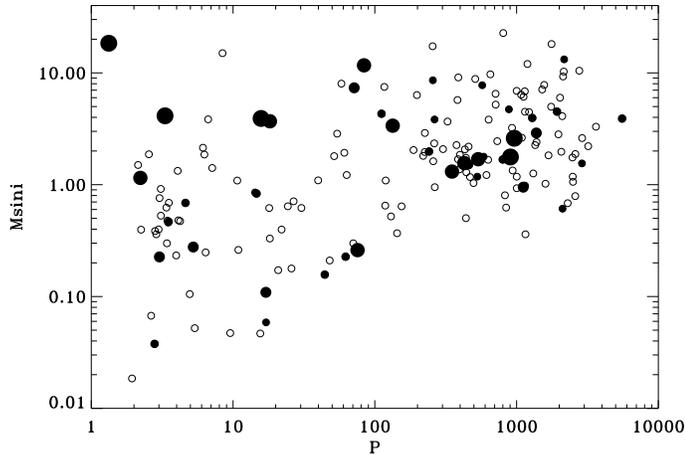}
      \caption{Projected mass vs orbital period of extrasolar planets.
               Open circles: single stars; filled circles: binary stars.
               The size of the symbol is proportional to the critical semimajor axis
               for dynamical stability (larger symbols refer to the tighter binaries).}        
         \label{f:msini}
   \end{figure}

The possibility that the Zucker \& Mazeh correlation was simply due to
the small number of planet in binaries known at that time can be considered.
However, an alternative interpretation appears more convincing when one
takes into account the binary properties of the massive hot-Jupiter hosts:
these are mostly relatively tight binaries (Fig.~\ref{f:msini}).
Indeed, Zucker \& Mazeh (\cite{zucker02}) considered in their statistical
analysis only binaries whose 
projected separation is smaller than 1000 AU, then excluding very wide
binaries included in increasingly larger number in later studies.

\begin{figure}
   \includegraphics[width=9cm]{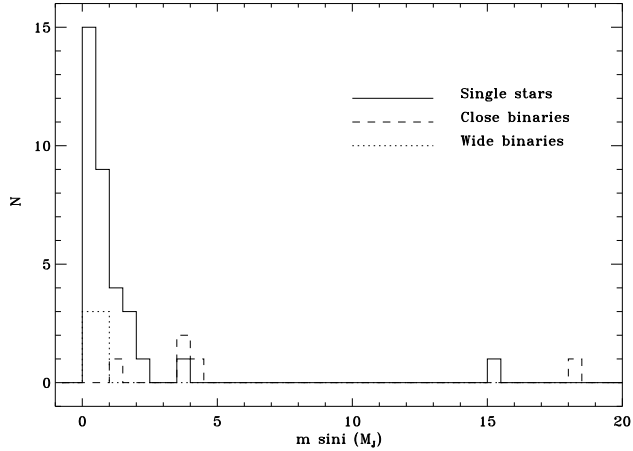}
   \includegraphics[width=9cm]{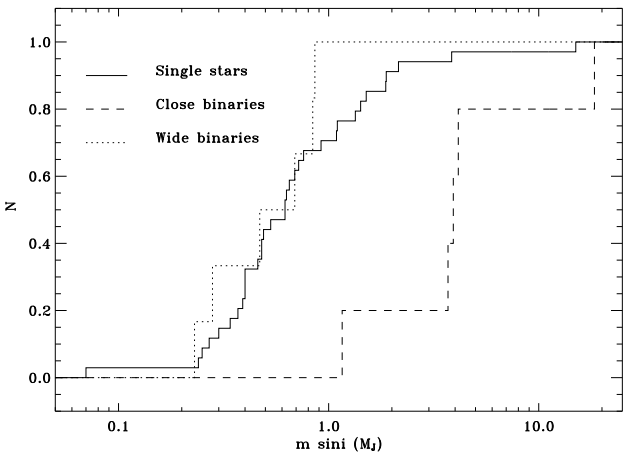}
      \caption{Distribution of the projected mass $m  \sin i$ for the 
               planets with period shorter than 40 days and $K<15$ m/s 
               orbiting single stars (continuous line), planets orbiting 
               binaries with $a_{crit} <75$~AU (dashed line), and planets 
               orbiting  binaries with $a_{crit} > 75$~AU (dotted line). 
               Upper panel:
               histogram; lower panel: cumulative distribution.}
         \label{f:hist_51peg}
   \end{figure}

To test this hypothesis statistically, we selected from
Table \ref{t:tablebin} and B06 the planets with
$P<40$~days  (about in the middle of the so-called 'period
valley' on the period distribution of planets, see
Udry et al.~\cite{udry03} and Fig.~\ref{f:hist_p_all}).
We performed a KS test comparing
the mass distribution of planets with $P<40$~days orbiting  
single stars, binaries with  $a_{crit}<$ 75 AU and
binaries with $a_{crit}>$ 75 AU (34, 5, and 6 planets respectively)
(Fig.~\ref{f:hist_51peg}).
The hypothesis that the mass of close-in planets in tight
binaries can be drawn from the same parent distribution of close-in planets
orbiting members of wide binaries or single stars can be rejected
with 98.2\% and 99.5\%.
The MWU test yields an even higher significance ($>99.9$\%).
The mass distribution of close-in planets orbiting single stars and 
wide binaries  is not statistically different. 

We checked the effects of the assumptions made on our analysis 
on the resulting significance.
The significance of the difference of the mass distribution between
planets in tight binaries and single stars is larger than 99.9\% for
 $a_{crit}$=50 AU, it is about 97.5\%  for $a_{crit}$=100 AU and  
becomes smaller moving the limit  beyond 100 AU.
The exclusion of super-planets more massive than $10~M_{J}$
has some effect but the significance remains high ($>99$\%) for both 
 tests. Changing the period limit of close-in planets 
in the range 20-200 days keeps the significance always larger than 99\%.
The  RV semi-amplitude selection limit plays little role.

For some of the close-in planets with $m \sin i > 1.5~M_{J}$
orbiting stars classified as single there are indications of the presence
of significant dynamical perturbations.
\object{HD 118203 b}, \object{HD 68988 b} and \object{HIP 14810 b} have an 
unusually high eccentricity 
($e=0.31$, $0.14$, and $0.15$ respectively) for their orbital 
period ($\sim 6$ days). 
A second planet is orbiting \object{HIP 14810} (B06), while in 
the other two cases 
linear trends in the residuals from the short period orbit indicate 
the presence of a further companion in a long period orbit. 
Their mass and period will be revealed by the continuation of the
observations, allowing a proper classification of their hosts as binary systems
or multi-planet hosts.
A candidate companion close to \object{HD 162020} was reported by 
Chauvin et al.~(\cite{chauvin06}), whose physical association remains to be established.
For \object{HD 73256 b} there is currently no indication
of additional companions. 

\subsection{The mass distribution of planets in wide orbits}
\label{s:wide}

\begin{figure}
   \includegraphics[width=9cm]{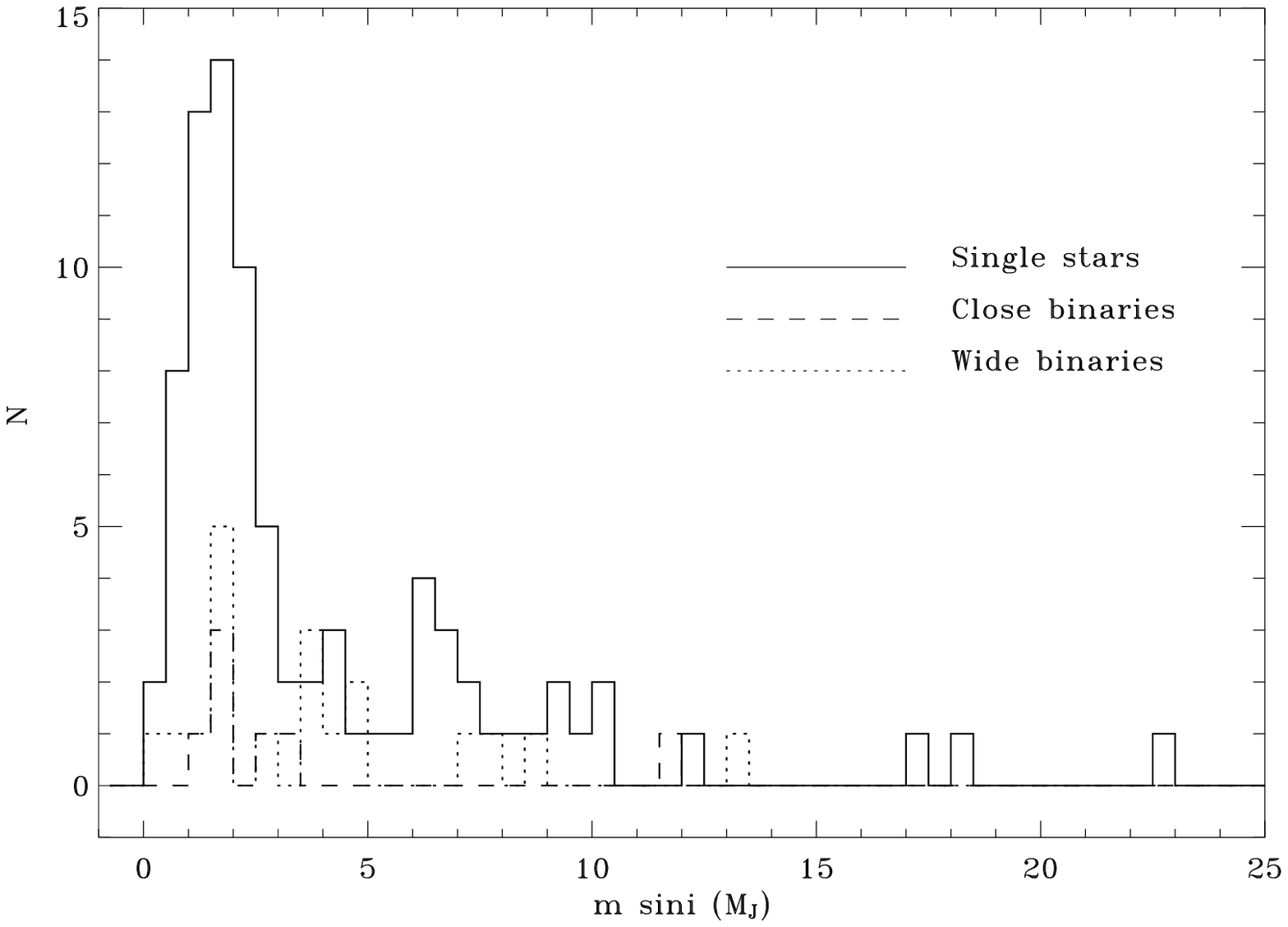}
   \includegraphics[width=9cm]{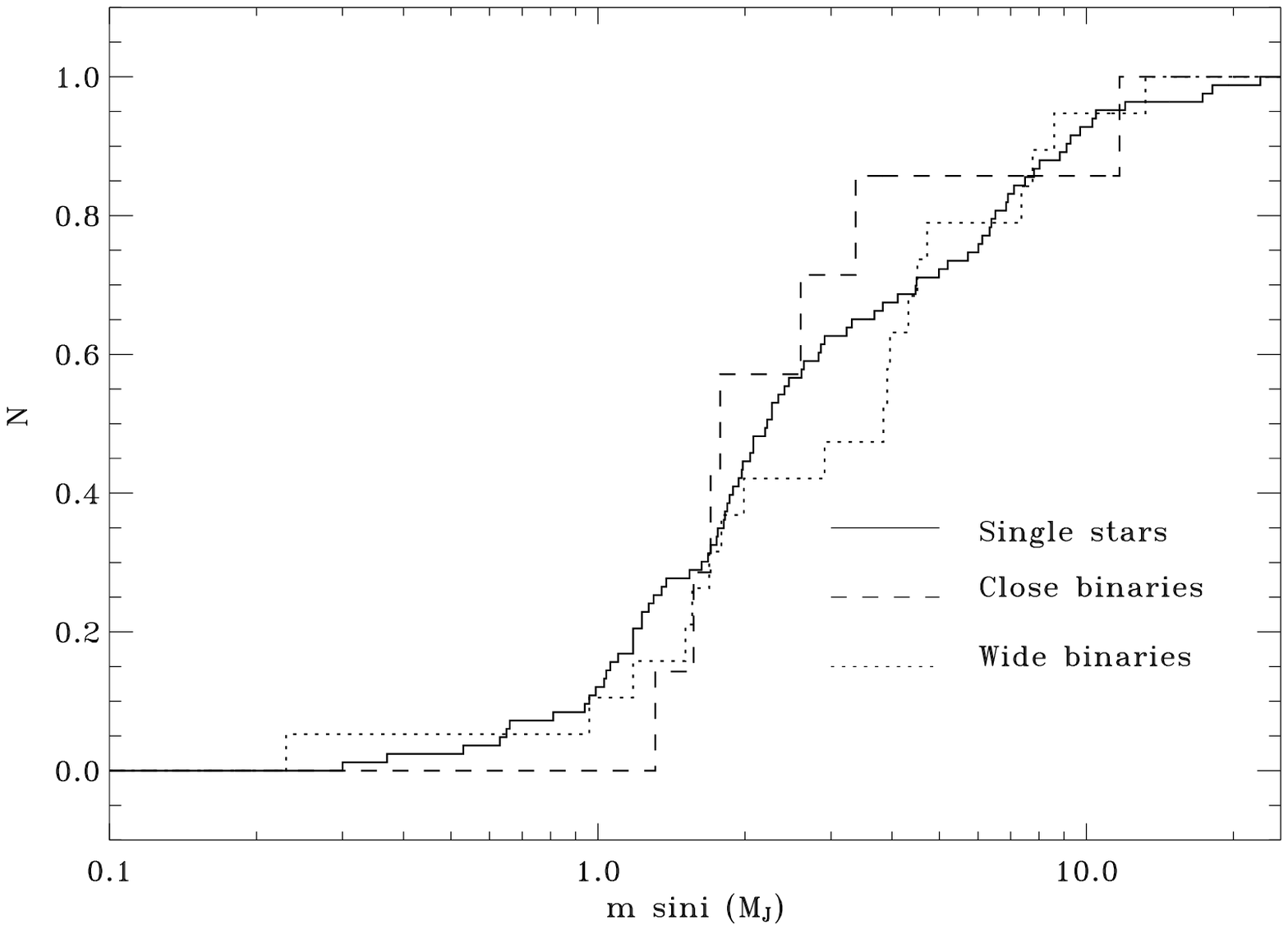}
      \caption{Distribution of the projected mass $m  \sin i$ for the planets with period
               longer than 40 days orbiting single stars (continous line), planets orbiting binaries with
               $a_{crit} <75$~AU (dashed line), and planets orbiting 
               binaries with $a_{crit} >75$~AU (dotted line). Upper panel: histogram; lower panel: cumulative
               distribution.}
         \label{f:hist_long}
   \end{figure}

Fig.~\ref{f:hist_long} shows the distribution of $m \sin i$
of planets with $P> 40$ days.
The lack of massive planets with $P> 100$ days
in binaries noted by E04 is no longer valid: three planets with $m \sin i \ge 5.0~M_{J}$
are now in the sample. All of them are orbiting the components of very
wide binaries (projected separation larger than 800 AU).
The probability of having 3 planets with $P>100$ days  and   $m \sin i \ge 5.0~M_{J}$
derived using the hypergeometric distribution is 17\% when considering only 
wide binaries and 7.4\% when considering all the binaries together. 
However, this probability becomes larger than 10\% when considering 
a different mass limit ($4-6~M_{J}$) and KS and MWU tests do not
reveal statistically significant differences. 

The mass distribution of long-period planets around tight binaries
(7 planets)
is not statistically significantly different with respect
to that of planets in single stars and wide binaries (83 and 19 
planets respectively).

\subsection{The period distribution}
\label{s:period}

The period distribution of short period planets shows a pile-up for periods
around 3 days (see e.g. B06).
Only one of the five close-in planets in tight binaries has a period 
close to 3 days (\object{$\tau$ Boo b}).
Two have a very short period (\object{HD41004Bb} and \object{HD 189733b} 
have the shortest and the third shortest period 
in the B06 catalog), while \object{GL86b} and \object{HD 195019b} have 
longer periods (15 and 18 days respectively). 
The differences are not statistically significant using KS and MWU tests.

When considering the period distribution of planets with $P>40$ days, there 
are hints that the period distribution of planets in tight binaries is
different, with a lack of planets with periods longer than 1000 days.
However, the KS and MWU tests do not show statistically significant 
differences.

\begin{figure}
   \includegraphics[width=9cm]{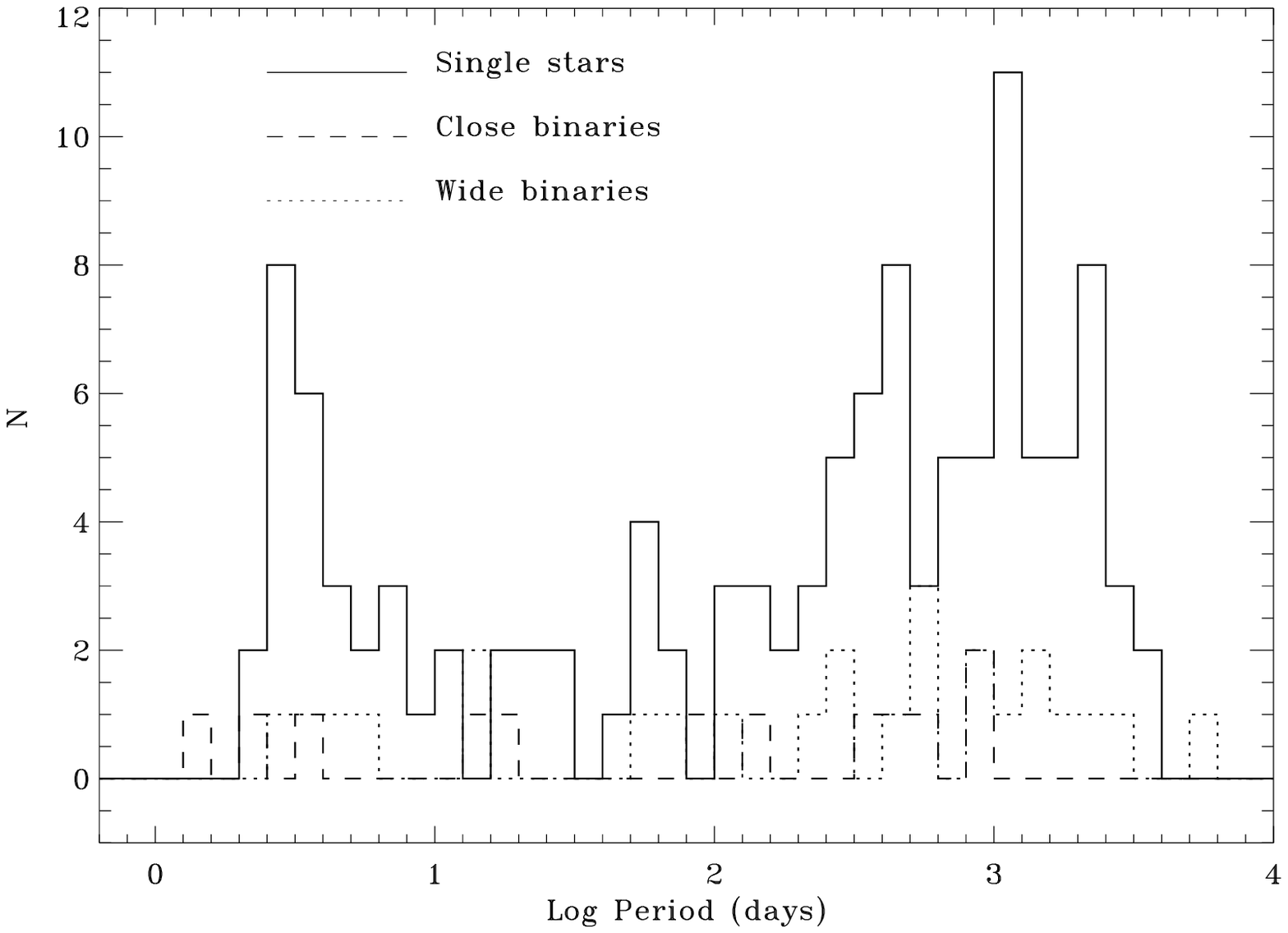}
   \includegraphics[width=9cm]{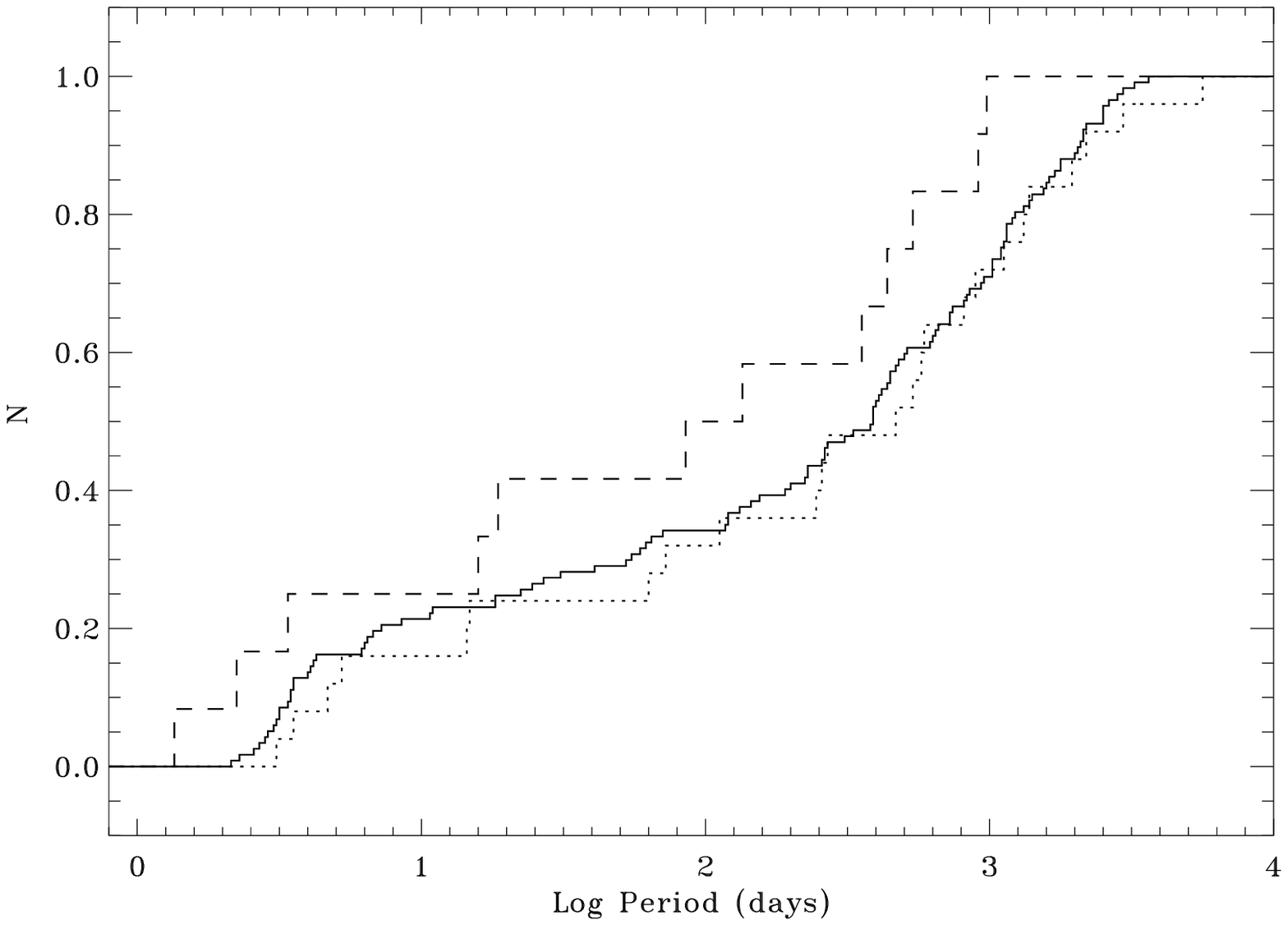}
      \caption{Distribution of log(period) for the planets 
               orbiting single stars (continous line), planets orbiting binaries with
               $a_{crit} <75$~AU (dashed line), and planets orbiting 
               binaries with $a_{crit} >75$~AU (dotted line). Upper panel:
               histogram; lower panel cumulative distribution.}
         \label{f:hist_p_all}
   \end{figure}

When considering the whole period range (Fig.~\ref{f:hist_p_all}),
the difference between the period distribution of planets around single stars 
and tight binaries (117 and 12 planets respectively)
is more significant (94.3\%) for the MWU test while it remains not significant
according to the KS (50\%). Similar values of significance are derived 
for the comparison between the period distribution of planets in tight vs
planets in wide binaries (25 planets).
The discrepant level of significance between the two tests can be understood
considering that the period distribution of planets in tight binaries 
appears to be shifted with respect to those orbiting single stars for 
the whole period range,
but without very large deviations. The maximum difference of the
cumulative distributions, the estimator used by the KS test,
is then not significant, while the global shift is
marginally significant by
the MWU-test, that is sensitive instead to differences of the median of the distributions. 

As the significance is not large and the different surveys have their 
own period sensitivity
and biases concerning binarity (see Sect.~\ref{s:bias}), 
we conclude that the evidence of the different period distribution
for planets orbiting single stars and components of tight binaries
needs confirmation.

The period distribution of planets  orbiting single stars and 
components of wide binaries are instead remarkably
similar considering both separate period ranges and for the whole
period range.

\subsection{The eccentricity distribution}
\label{s:ecc}

The possible link between high planet eccentricity and binarity has been 
considered since the discovery of the planet orbiting \object{16 Cyg B}
(Holman et al.~\cite{holman97}; 
Mazeh et al.~\cite{mazeh97}).
It is thus relevant to consider the eccentricity distribution
for planets orbiting tight and wide binaries and single stars.

\begin{figure}
   \includegraphics[width=9cm]{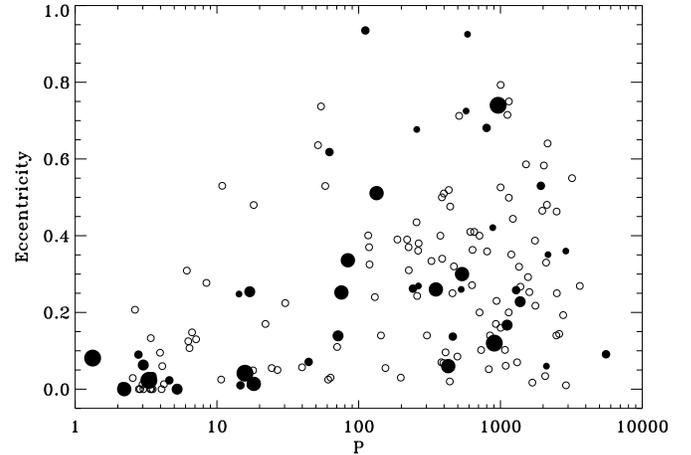}
      \caption{Eccentricity vs orbital period for planets in binaries (filled circles) and
               orbiting single stars (empty circles). Different sizes of filled circles refer 
               to different periastron of the binary orbit (larger sizes: closer orbits).}
         \label{f:p_e}
   \end{figure}

Fig~\ref{f:p_e} shows the period-eccentricity diagram.
The only two planets with $e>0.80$ (\object{HD 80606b} and \object{HD 20782b}) are 
in binary systems. However,  planets with $e$ up to 0.75 appear to
be present also around stars currently known as single.

\begin{figure}
   \includegraphics[width=9cm]{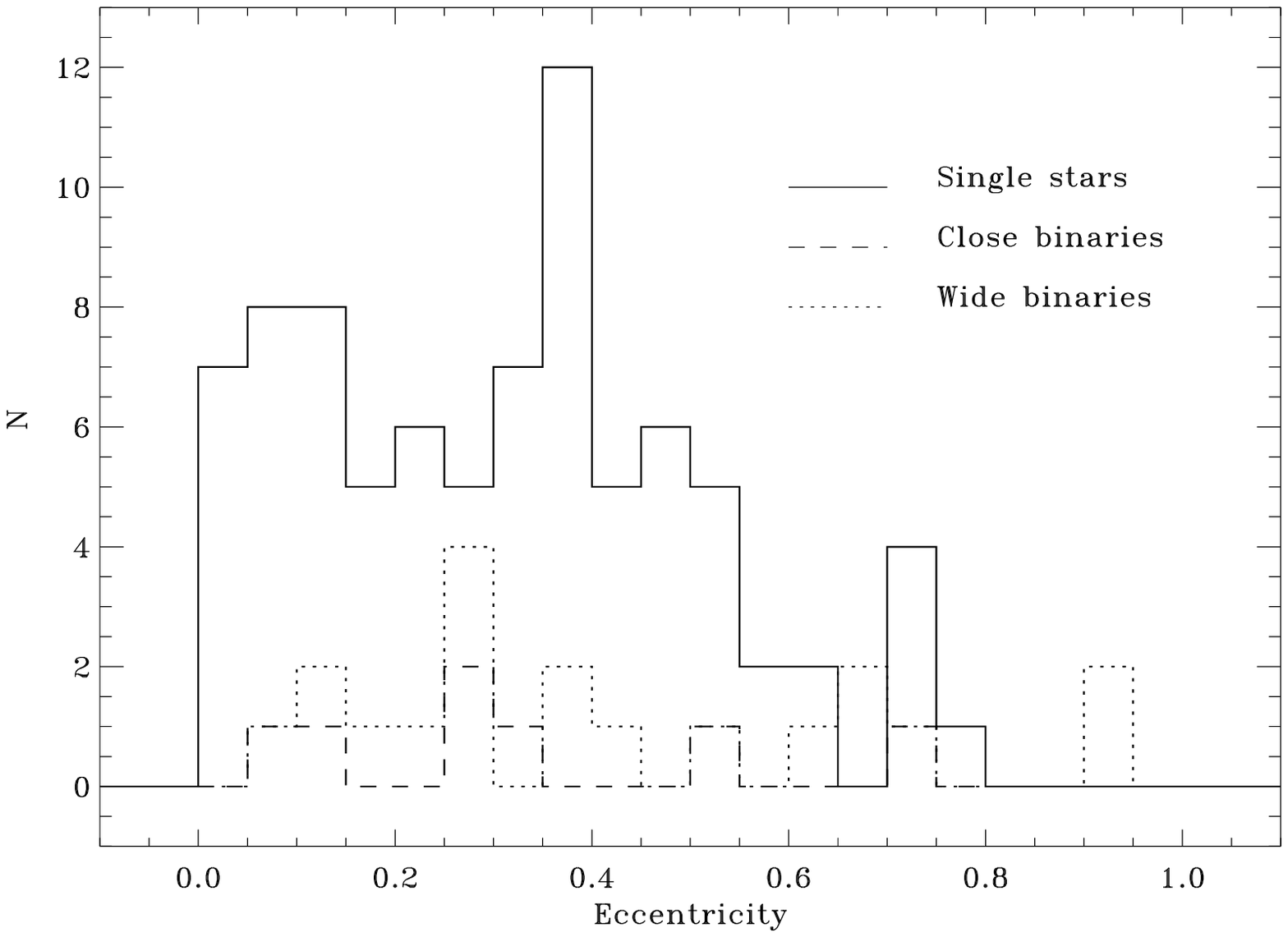}
   \includegraphics[width=9cm]{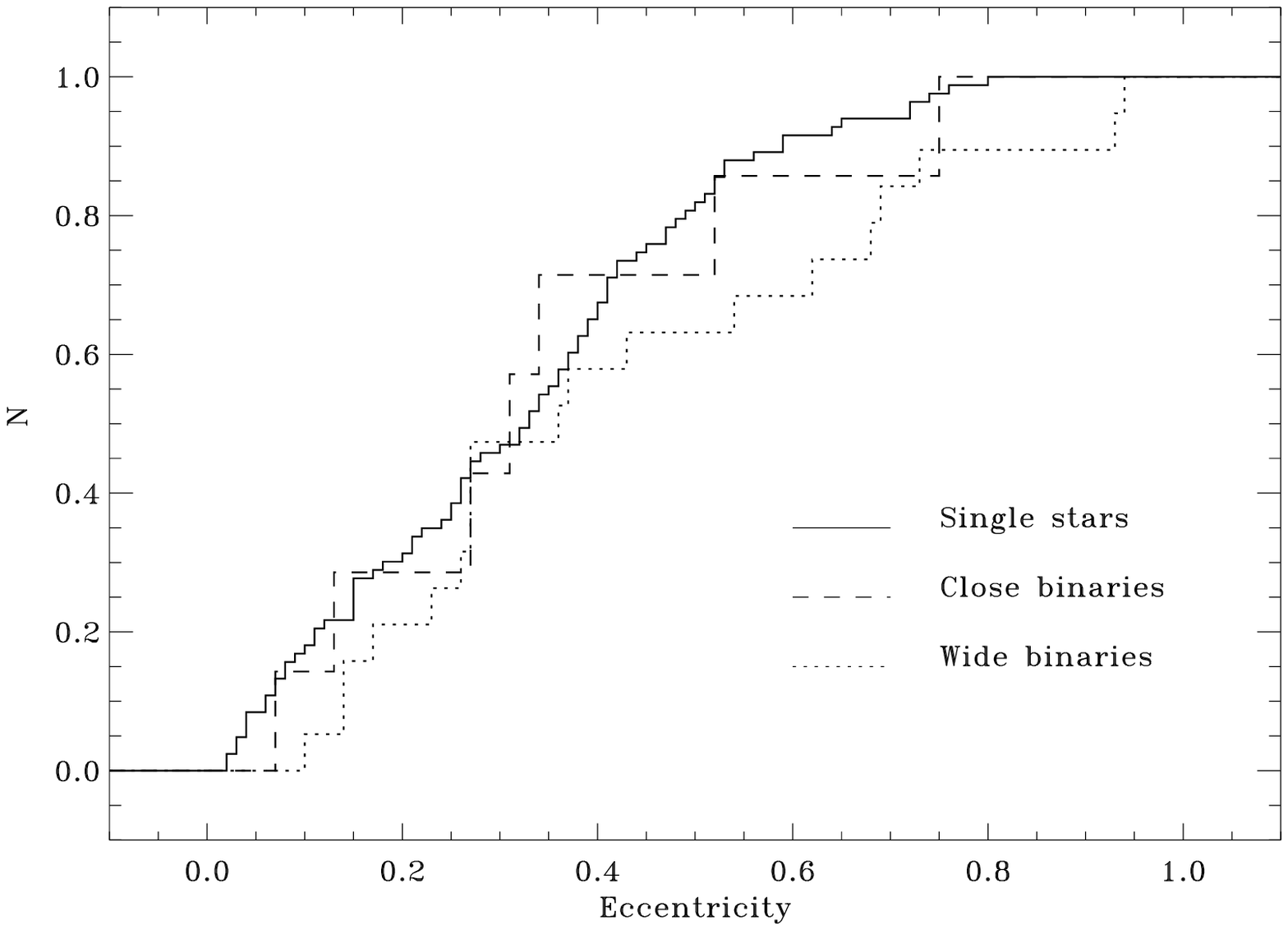}
      \caption{Cumulative distribution of the planet eccentricity for the planets with period
               longer than 40 days orbiting single stars (continous line), planets 
               orbiting binaries with $a_{crit}$ smaller than 75 AU (dashed line), 
               and planets orbiting binaries with $a_{crit}$ larger than 75 AU (dotted line).}
         \label{f:cumul_ecc}
   \end{figure}

The eccentricity distribution of planets in tight binaries with 
periods longer than 40 days (Fig.~\ref{f:cumul_ecc}) is not significantly 
different  to those
orbiting single stars\footnote{The large 
eccentricity of \object{HD41004Ba} is uncertain ($e=0.74\pm0.20$).}. 

The comparison for  planets in wide binaries  
indicates no significant differences according to the KS test (74\%), while 
the MWU test gives a significance of 91.2\% that the median of the two samples 
are different. 

E04 noted that all the five planets in binaries in their sample with periods shorter than
40 days have an
unusually low eccentricity ($e<0.05$). They estimated a chance probability 
of 3.8\% using the hypergeometric distribution.
While a tendency to a low eccentricity of short-period planets in binaries seems to be still
present (4 planets of 5 and 4 of 6 with  $e<0.05$, for the tight and wide
binaries respectively, compared with 17 of 34 for planets in single stars)
a similar analysis in our sample (taking together wide and tight binaries) yields 
a probability of 12\%, therefore not confirming the E04 result at a highly
significant level. KS and MWU tests also do not suggest statistically
significant differences.
Selecting only planets with periods  between 10 and 40 days (to exclude
objects whose orbits are circularizated by tidal effects) 
does not increase the significance
of a lower eccentricity for short-period planets in binaries.

\subsection{Multi-planet systems }
\label{s:systems}

The occurrence of systems with more than one  planet 
around the components of wide binaries 
(4 of 23, $17.3\pm12.3$\%) is similar with respect to that of planets 
orbiting single stars
(15 of 112, $13.3\pm4.7$\%).
On the other hand ,no multiple planets have been yet discovered around the 
components of tight binaries (13 objects). 
The closest binary with a known multi-planet system is 
$\upsilon$ And (projected separation 750 AU).
However, the small number of tight binaries in the sample
makes the lack of multi-planet systems
not highly significant (probability of 15\% of occurring by chance using
the hypergeometric distribution).
The similar fraction of multiplanet systems around single stars and
wide companions is a further indication that the presence of a wide companion
does not alter too much the process of planet formation.

\subsection{Metallicity of planets in binary systems}
\label{s:metal}

\begin{figure}
   \includegraphics[width=9cm]{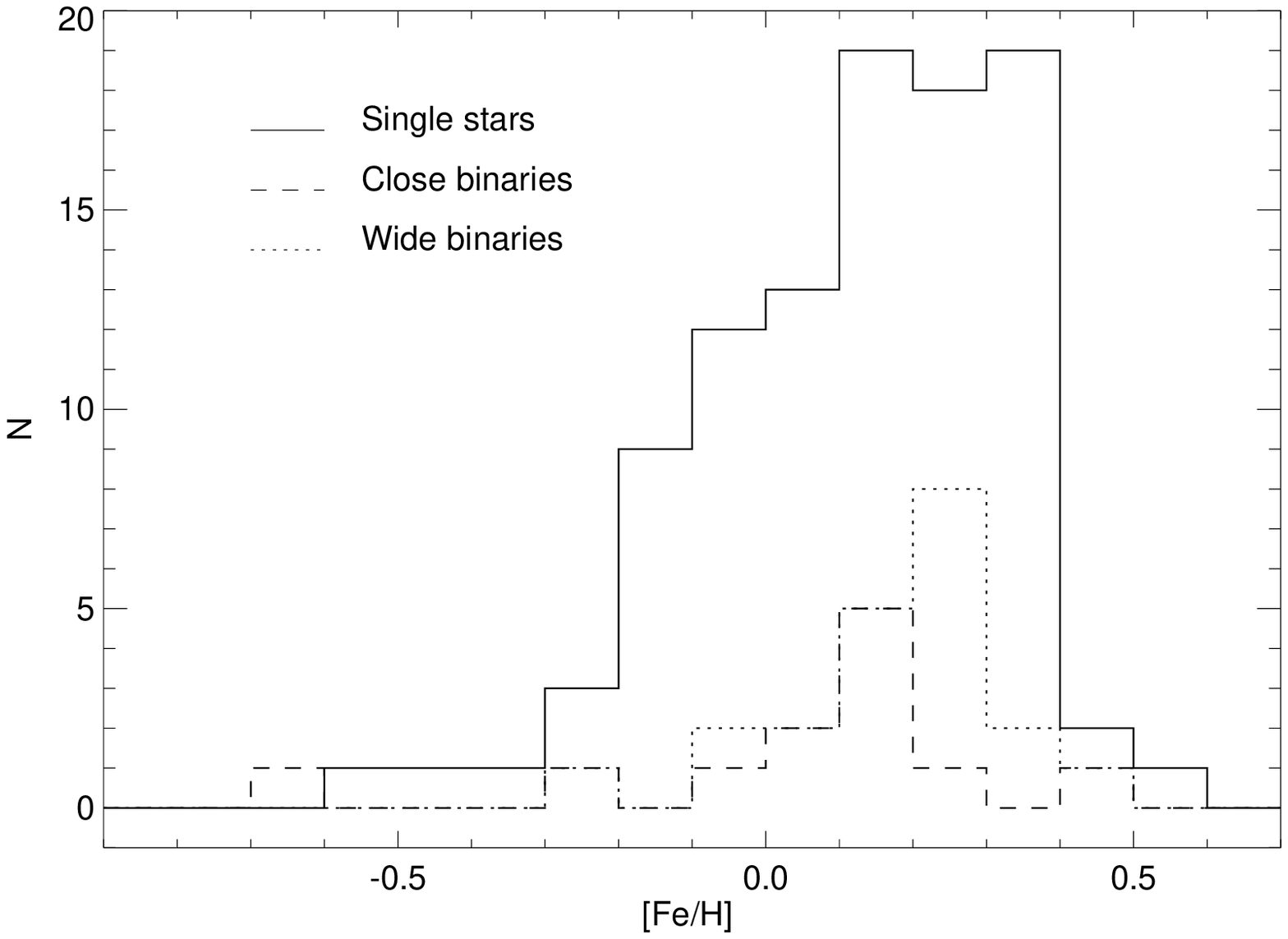}
   \includegraphics[width=9cm]{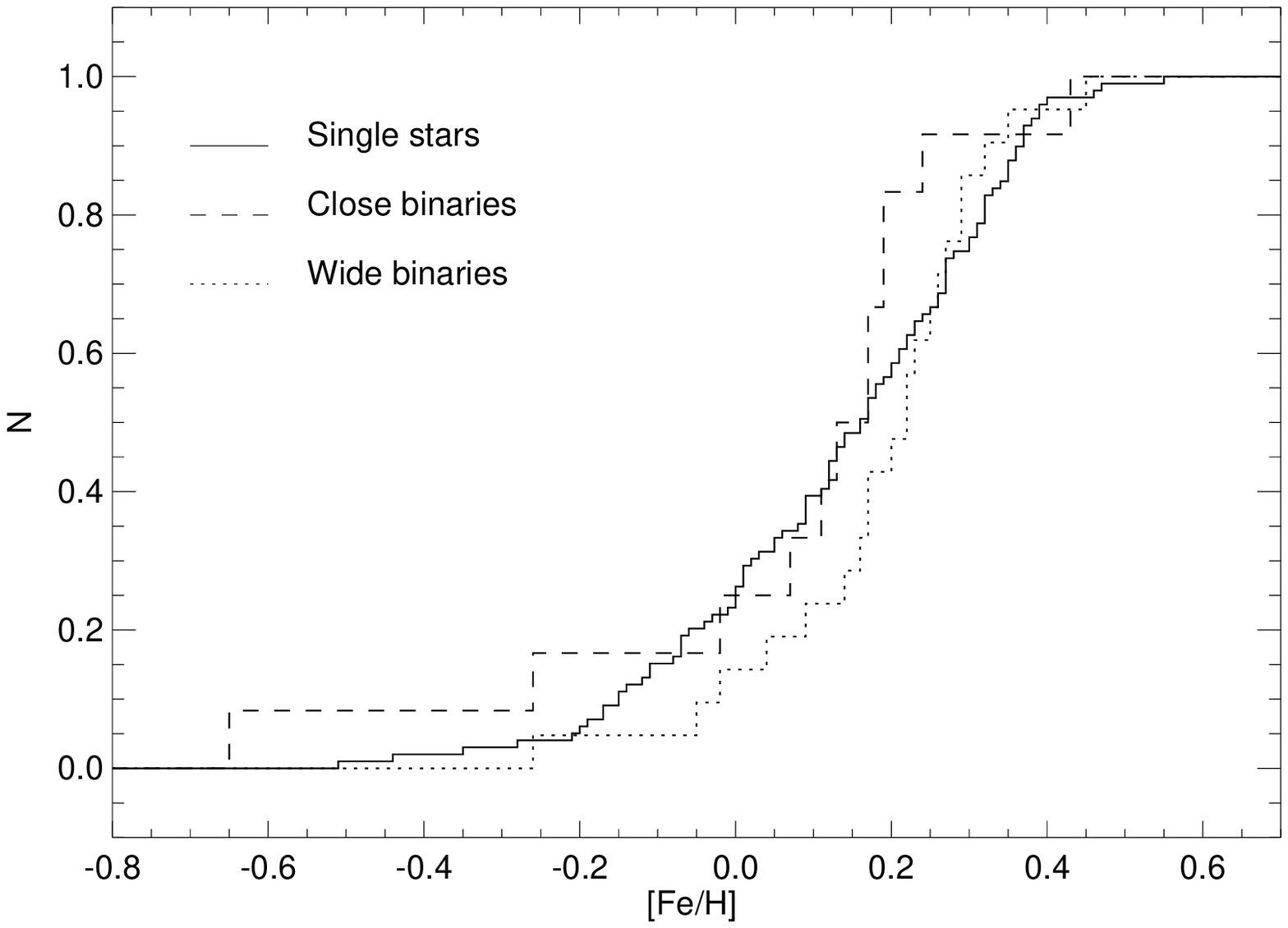}
      \caption{Distribution of the stellar metallicity [Fe/H] for the planets 
               orbiting single stars (continous line), planets orbiting binaries with
               $a_{crit}<75$ AU (dashed line), and planets orbiting 
               binaries with $a_{crit}>75$ AU (dotted line).
               Upper panel: histogram; Lower panel: cumulative distribution.}
         \label{f:hist_feh}
   \end{figure}

The study of the metallicity of planet host in binaries
can shed light on possible different planet formation mechanisms
(see Sect.~\ref{s:discussion}).
The close-binary planet-hosts appears slightly more metal poor
than single stars, while the wide-binary planet-hosts are slightly
more metal rich.
The comparison of cumulative distributions (Fig.~\ref{f:hist_feh}) 
using KS and MWU  tests indicates  probabilities of 
12.1\% and 4.6\% respectively that the metallicity
of close and wide binary planet hosts can be drawn from the same
parent population.
The differences with respect to single stars are less significant
(probability of 23.3\% and 17.8\% for the close binary vs single stars and
and wide binary vs single stars respectively using the MWU test, a
smaller significance  using KS).
Considering together all the binaries makes any difference not significant.

While intriguing, the possible differences of the metallicity of binary
planet hosts should be taken with caution.
Beside the rather large probability of a chance result, 
it should be considered that the small number
of planets in close binaries  makes  the role
of individual objects critical (e.g. \object{HD 114762}, whose
companion might be a brown dwarf or even a low mass star,
Cochran et al.~\cite{cochran91}).
Furthermore, indirect selection effects might
cause spurious correlations (see Sect.~\ref{s:bias}).

\subsection{Selection effects and caveats}
\label{s:bias}

The B06 catalog includes planets discovered by several different surveys.
Different RV surveys
typically have their own biases concerning binarity and other
properties (e.g stellar mass and evolutionary phase, metallicity, etc.).
Furthermore, they have different precision and time spans, and thus
different planet mass and period sensitivities.

All these  effects cannot  controlled in our study, and 
we cannot exclude that they play a role in producing some of the
marginal differences we showed (e.g. in  metallicity 
and the period distributions).
However,  the most significant result, the different mass
distribution of close-in planets in tight binaries, should not be affected
by these selection effects.
In fact, these planets are the easiest to discover for any survey
because of their
short periods and large radial velocity amplitude.

An additional caveat is represented by the uncertainty in the
upper limit of planetary mass, which is assumed in this study
to be $24~M_J$ following B06, and by the other choices performed 
in the analysis (e.g. period limits, etc.)

The sample of binaries among planet hosts
is certainly not complete. A large fraction of the planet hosts
were scrutinized only for comoving companions at rather large
separations (a few arcsec).
The analysis was performed considering
the actual binary configuration that in some cases can be significantly
different  to that at the time of the formation of the
planets. In the case of binary systems including a white dwarf,
the probable original configuration can be derived estimating the original
mass of the (now) white dwarf (see Appendix~\ref{s:wd}), but  other
cases are speculative (see Appendix\ref{s:hd20782}).


\section{Discussion}
\label{s:discussion}

Exploiting the continuously expanding samples
of extrasolar planets, we investigated the
occurrence of a difference between planets in binaries
and planets orbiting single stars.
The size of the sample allowed us to divide the 
binaries in two sub-groups, 'tight' and 'wide' binaries.
The separation was fixed at a critical semiaxis for dynamical
stability equal to 75 AU. This corresponds to separations of
about 200-300 AU depending on the mass ratio between
the components.

\subsection{The properties of planets in tight binaries}
\label{s:tight}

The most significant feature we were able to show is that
the mass distribution of close-in planets around 
close and wide binaries is different.
Only the close-in planets in tight  binaries
have a mass distribution significantly different to those
orbiting single stars.
This result  implies that the formation and/or migration and/or 
dynamical evolution processes acting in the presence
of a sufficiently close external perturber are modified with respect to 
single stars.

The fact that the hosts of massive planets in close orbit are typically
binaries somewhat resembles
the results by Tokovinin et al.~(\cite{tok06}) i.e. short-period
spectroscopic binaries have in most cases a further companion.
That a third member could act as an angular momentum sink 
resulting in a closer inner binary
was first suggested by Mazeh \& Shaham (\cite{mazeh79}). 
Tokovinin et al.~(\cite{tok06}) favor generation 
of high orbital
eccentricity by Kozai resonances, and subsequent tidal circularization,
as the mechanism to explain the close spectroscopic binaries
and their link with the presence of additional companions.

While it cannot be excluded that some of the massive close-in planets
in binaries were moved into their current location by Kozai
migration, some other massive planets (\object{GL 86b} and \object{HD 195019b})
have periastrons of 0.108 and 0.137 AU, probably
excluding the occurrence of tidal circularization.
This calls for a more general process to explain
the presence of massive close-in planets.

A possible effect on migration and mass accumulation 
timescales in a disk in the presence
of a companion
was  predicted by the models by Kley (\cite{kley00}).
He considers the evolution of a Jupiter mass planet embedded in
a circumstellar disk in  binary
systems with a separation of 50-100 AU, with eccentricity $e_{bin}=0.5$,
and stellar masses of 1.0 and 0.5 $M_{\odot}$. 
The presence of the companion significantly  enhances the growth rate
and make the migration timescale of the planet shorter.
As discussed by E04, this can  qualitatively explain the
presence of massive close-in planets in binary systems.
A step forward here is that we have shown
that most of the binary systems that host the 'anomalous' massive,
close-in planets are indeed not too  different 
to those modeled by Kley (\cite{kley00}), while E04 were
forced to consider  also binary systems with very
wide separation, for which the described model is likely not to work.
The models by Kley (\cite{kley00})
have as an initial condition an already formed planet and thus
do not include the  formation phase of giant planets.
As discussed by Thebault et al.~(\cite{thebault06}), 
a possible way to achieve the formation of planets in a swarm
of planetesimals perturbed by a stellar companion might be
type II runaway accretion identified by 
Kortenkamp et al.~(\cite{kortenkamp01}).
More detailed modeling is  required to achieve a full understanding
of the processes that cause the differences of planet
properties in tight binaries.

Additional possible differences of the properties of planets in tight binaries 
(to be confirmed as the statistical significance is not very high
and selection effects might be at work) are
the different period distributions (there are
no planets with $P>1000$, the close-in planets probably do
not pile-up at period close to three days) and the 
lack of multi-planet systems.

In a few cases,
the lack of long period planets can be understood 
as the result of  dynamical stability constrains: for 
\object{$\gamma$ Cep} and \object{HD 41004A}, the critical semimajor axis
for dynamical stability is just 1.9 and $\sim$3 times the planet semimajor 
axis.
For the other cases, a more general mechanism should be at work
(enhanced migration; Kley \cite{kley00} or  lack of suitable conditions for 
planet formation in the
external parts of the disks, under the gravitational 
influence of the companion).

The fact that properties of planets in tight binaries 
are different  to those orbiting
single stars can be exploited to distinguish between different
scenarios for their origin.
As discussed by e.g. Hatzes \& Wurchterl (\cite{hatzes05}) 
and Thebault et al.~(\cite{thebault04}), the presence of planets
in binaries with a separation of 20 AU or less represents
a challenge for our current view of planet formation.
Such problems can be overcome if the binary was initially
wider, and its orbit was modified  through dynamical
interactions in star clusters after planet formation
(Pfahl \& Muterspaugh \cite{pfahl06}).
The different mass distribution of planets in close binaries
implies that, if this is the case, 
the dynamical interactions should 
not only modify the binary orbit but also
force a massive planet in external orbit to migrate inward, as
massive hot Jupiters are rare around single stars.
This point should be addressed by future dynamical simulations.
Kozai oscillations and tidal circularization after the dynamical
interaction cannot be excluded in some cases
(large relative inclinations between the planet and binary
orbits might  be the result of the proposed  dynamical encounters).
However, less massive planets originally
in external orbits should follow a similar fate and thus we expect
that low mass planets should also be present in close orbits 
in tight binaries, at odds with current observational data.
Furthermore, as discussed above, Kozai migration cannot
explain planets with periods longer than a few days
and low eccentricities.

Further clues to the origin of planets in tight binaries 
can be derived by a determination of their frequency.
Pfahl \& Muterspaugh (\cite{pfahl06}) predict that the fraction
of close binaries\footnote{The definition of close
binaries used by Pfahl \& Muterspaugh (\cite{pfahl06}) (semimajor
axis less than 50 AU) is different  to that
adopted here.} that dynamically acquire giant planets 
is about  0.1\%, with an uncertainty of about one order
of magnitude.
They note that the rough number of these planets seems
too high to be compatible with these predictions.
However, the number of tight binaries
included in the radial velocity samples is not well known.
A more detailed analysis is postponed to a forthcoming study.

The mentioned difficulties of the binary-interaction scenario
to account for the occurrence of planets in close binaries
leave open the alternative hypothesis, i.e.
giant planets formed in binaries with small
separation at the time of planet formation, 
possibly in a different way than planets around single stars.

The metallicity distribution of tight binaries with planets
can give clues to the formation mechanism for these planets.
The large effect of the stellar metallicity on the frequency
of planetary companions is well established (Fischer \& Valenti \cite{fv05};
Santos et al.~\cite{santos04}) and can be understood in the
framework of the core-accretion model for the formation of giant
planets (Ida \& Lin \cite{ida04}; Robinson et al.~\cite{robinson06}).
If these planets formed in a different way than those
orbiting single stars and the components of wide binaries, e.g.
by disk instability triggered by dynamical perturbations (Boss \cite{boss06}), 
we should expect a lower metallicity for planet hosts in close
binaries, as the disk instability mechanism is expected to be
rather insensitive to metallicity (Boss \cite{boss02}).
There is a marginal indication 
of a larger fraction of metal poor stars among
tight binaries with planets. However, the distribution of 
metallicity remains 
on average larger than that if the solar vicinity, with  a mean
and median larger than the solar value.
This indicates that high metallicity is a factor favoring planet formation 
also in tight binary systems.

\subsection{The properties of planets in wide binaries}
\label{s:wide}

The mass and period distributions of planets in wide binaries
are not statistically significant different 
to those of planets orbiting single stars.
The lack of long-period massive planets in binaries shown by E04 
is not confirmed
at a highly significant level in our study.
The fraction of multi-planet systems is similar for planets orbiting
single stars and components of multiple systems.
These results indicate that a distant companion (separation $>$300-500 AU)
does not  significantly affect the process of planet formation.
The only marginally significant difference between planets orbiting single
stars and components of wide binaries concerns the planet eccentricity.

\subsection{Is binarity the cause of the high eccentricities ?}
\label{s:ecc}

The eccentricity distribution of planets in wide binaries shows
a marginal excess of high-eccentricity ($e \ge 0.6$) planets.
However, current results indicate that the high planet eccentricity is not 
confined to planets in binaries,
and that the possible differences in eccentricity are limited
to the range $e \ge 0.5-0.6$.
This indicates that there are mechanism(s) generating planet 
eccentricity up to 0.4-0.5 that are independent of the binarity
of the planet host, and are characteristic of formation and
evolution of a planetary system
(e.g. disk-planet interactions: Tremaine \& Zakamska \cite{zakamska04};
 planet-planet scattering:
Marzari \& Weidenschilling \cite{marzari02},
Ford et al.~\cite{ford03}).
These probably act
during or shortly after planet formation.
Further eccentricity enhancements, possibly linked to the presence
of a companion, might take place
at later epochs. 
In fact, Takeda et al.~(\cite{takeda06}) noted that most  
high-eccentricity planets orbit old stars (ages $>$5 Gyr).  
Mechanisms that require long time scales to modify 
planetary orbits then seem favored.

A popular mechanism that might produce high eccentricity planets in binaries
is represented by Kozai oscillation.
Kozai oscillation is expected to be at work also for
low mass companions at large separation (the case of
most of the companions of planet hosts discovered up to now), 
provided that a sufficiently long time is available 
and that the planet has a sufficiently long period
to prevent the suppression of Kozai oscillation
by general relativistic effects
(Takeda \& Rasio \cite{takeda05}; Wu \& Murray \cite{wu03}).
As an example, a brown dwarf companion such as \object{HD 3651B}
(mass about $0.06 M_{\odot}$, projected separation about 500 AU)
might be able to induce the eccentricity oscillation for
a planet with an initial semimajor axis larger than about 3 AU. 

Kozai oscillations are expected to produce large planet
eccentricities.
Wu \& Murray (\cite{wu03}) considered in detail the possibility that
the high eccentricity of \object{HD 80606b} ($e=0.93$) is due to Kozai
oscillation. This was found to be possible only for a relative inclination
of the planetary and binary orbit larger than 85 deg.
In the case of the only other planet with $e>0.8$ (\object{HD20782b}), 
the period of the Kozai
eccentricity oscillation derived from Eq.~2 of  
Takeda \& Rasio (\cite{takeda05})
is much longer than the Hubble time.
Therefore, the extremely high eccentricities 
of  \object{HD20782b} is unlikely to be due to the
Kozai mechanism (see Appendix \ref{s:hd20782}), unless
further companions exist in the system at smaller
separation.

An alternative to  Kozai oscillations is represented by chaotic evolution
of planetary orbits induced by dynamical perturbations
(Benest \& Gonczi \cite{benest03}; Marzari et al.~\cite{marzari05}).
This kind of evolution might also arise for the 
dynamical interactions within the planetary system, without
requiring a stellar companion in a wide orbit.

\begin{figure}
   \includegraphics[width=9cm]{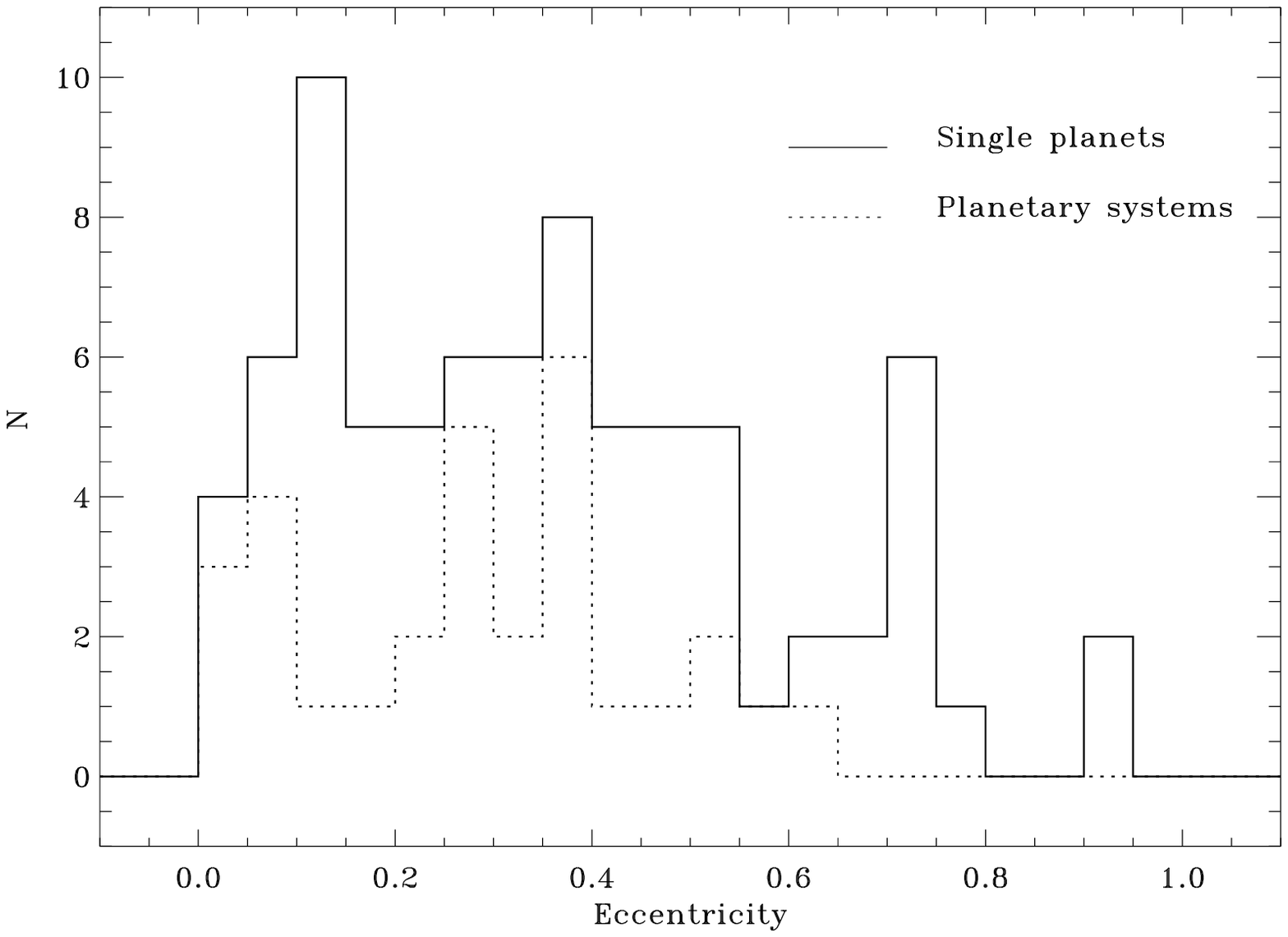}
   \includegraphics[width=9cm]{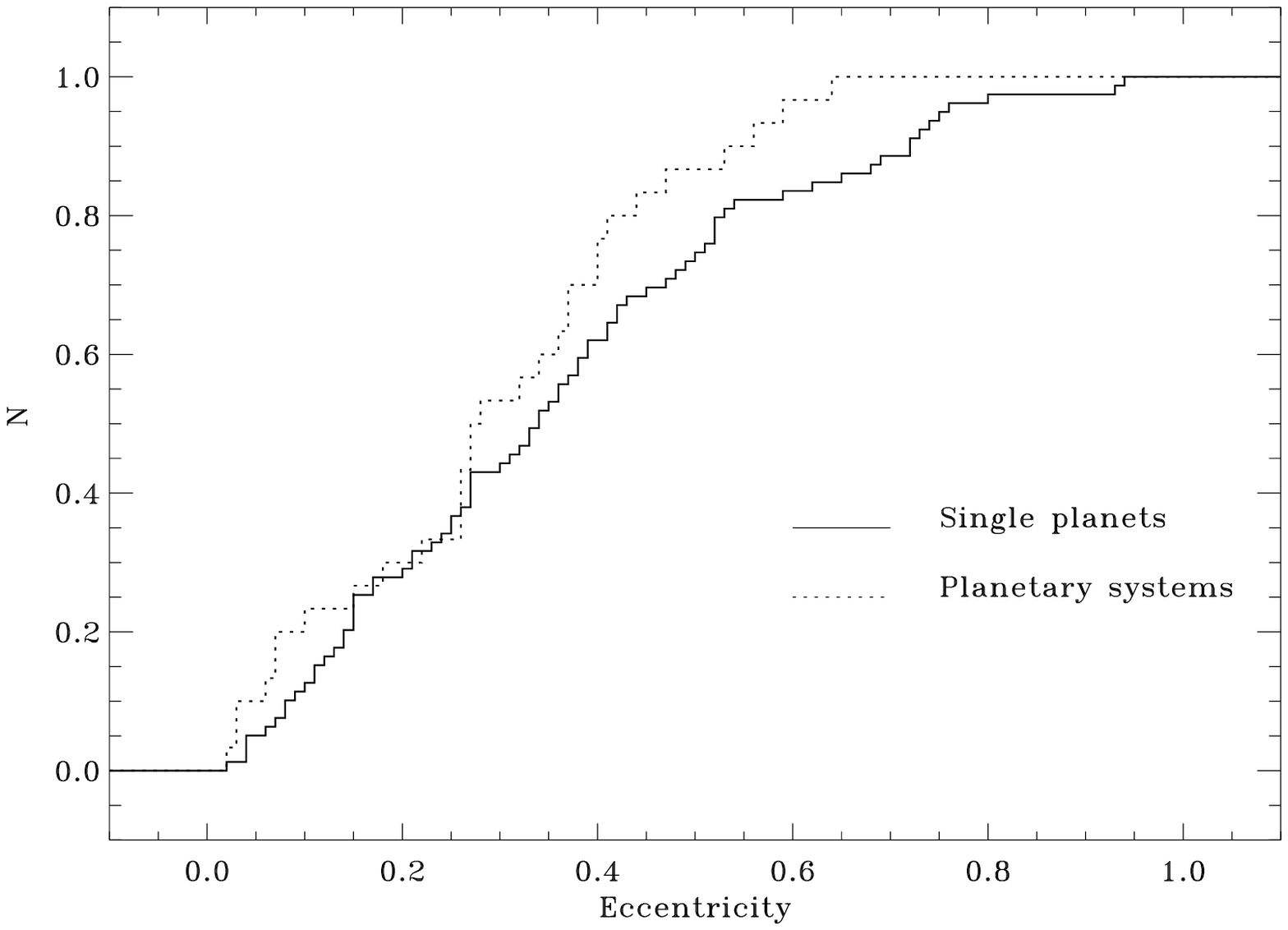}
      \caption{Distribution of the planet eccentricity for the single planets 
               (continous line), and planets in planetary systems (dotted line).
              Upper panel: histogram; lower panel: cumulative distribution.}
         \label{f:cumul_ecc2}
   \end{figure}

To further investigate the origin of the high eccentricities, 
we consider the eccentricity distributions
of planets residing in multiplanet systems and isolated planets 
(Fig.~\ref{f:cumul_ecc2}).
There is a lack of high eccentricity planets in multi-planet systems
(as noted also by Takeda et al.~\cite{takeda06}).
When considering planets with $P>40$ days,
the comparison of the two distributions reveals a difference that is 
significant
to about a 89.1\% confidence level (according to the MWU test), 
similar to the results of the 
planets in wide binaries vs those
in single stars. 
This can be understood considering that planetary systems with 
highly eccentric
planets are in general more subject to close encounters, and then
less stable, than a system populated with planets in low eccentricity
orbits. If the chaotic evolution of a planetary system has as a final
result a planet on a highly eccentricity orbit,
it is probable that that planet remains alone.
On the other hand, Kozai oscillations are more likely to take place
in single-planet systems, as they are suppressed in the presence of significant
mutual planet-planet interactions. Therefore, eccentricity enhancements
due to Kozai oscillation are expected to occur for single-planets
systems with a companion in a suitable binary configuration.
A proper evaluation of the link between high eccentricities, binarity of
planet hosts, and planet multiplicity requires larger samples to allow
more statistically significant inferences.

\section{Conclusion}
\label{s:conclusion}

The only highly significant difference shown in our analysis
concerns the mass distribution of short-period planets in tight
binaries.
Massive planets in short period orbit are found in most cases around
the components of rather tight binaries.
Other possible peculiar features of planets in tight binaries with
respect to planets orbiting single stars, such as a lack of long-period
planets and multiple planets, are of marginal significance and 
need confirmation.

The properties of exoplanets orbiting the components of wide
binaries are compatible with those of planets orbiting single stars,
except for a possible greater abundance of  planets
on highly eccentric orbits.
The previously suggested lack of massive planets with $P>100$ days in binaries 
is not confirmed.

This result indicates that the binary separation
plays a role in affecting the properties of planets 
and that the simple classification of planets in binaries vs planets
orbiting single stars is not adequate to describe the dynamical effects.
This opens the perspective for more detailed modeling 
of the role of stellar companions on the formation and evolution of
planetary systems.



\begin{appendix}


\section{Binaries among new planet hosts}
\label{s:newbin}

As planet announcements arise quite frequently, 
existing compilations quickly become incomplete.
Twenty-one  stars are not considered in the latest 
study of multiplicity of planet hosts 
(R06). Most of them are 
new planet hosts while some are stars with companions
with mass between 13 to 24 $M_{J}$, not considered by R06
but included in B06 catalog.

We searched for companions of these stars in existing
astronomical catalogs such as Hipparcos (ESA \cite{hipparcos}), 
CCDM (Dommaget \& Nys \cite{ccdm}), 
WDS (Mason et al.~\cite{wds}), and ADS (Aitken \cite{aitken}).
For the following 16 stars there are no evidence of binarity in the literature:
\object{HD 4308},
\object{HIP 14810},  
\object{HD 33283},   
\object{HD 66428},   
\object{HD 81040},   
\object{HD 86081},   
\object{HD 99109},   
\object{HD 102195},  
\object{HD 107148},
\object{HD 109143},
\object{HD 118203},
\object{GJ 581},
\object{HD 137510},
\object{HD 187085},
\object{HD 212301},
\object{HD 224693}. 

\object{HD 33564} and \object{HD 164922} are included in CCDM and WDS
(the latter also in ADS) as triple systems, but the
companions display different proper motion  to
the planet hosts and thus
they are not physical companions.
Two additional objects (\object{HD 20782} and \object{HD 109749}) 
are  binaries. The physical association of the companions
and their physical properties are discussed in the next subsections.
Finally, the binarity of \object{HD 189733} is discussed in a dedicated 
paper by Bakos et al.~(\cite{bakos06}).

\subsection{HD 20782}
\label{s:hd20782}

\object{HD 20782} is listed as a binary in the CCDM catalog 
 (CCDM 03201-2850).
The companion is \object{HD 20781}. The very large separation, 252 arcsec,
that corresponds to 9080 AU at the distance of \object{HD 20782} (36 pc) 
makes a detailed check of physical association mandatory. 
Table \ref{t:hd20782} lists the stellar parameters of the two 
stars. Hipparcos parallaxes and proper motions, and 
Nordstrom et al.~(\cite{nordstrom04}) radial velocities, are 
fully compatible within
errors for the two components. This strongly suggests
physical association.
Therefore,  \object{HD 20782} and \object{HD 20781}
form a very wide common proper motion pair. 
Considering the nominal radial velocity and proper motion
differences, the pair could be bound.
The observed  separation is very large but not  extreme for binaries
in the solar neighborhood (Poveda \& Allen \cite{poveda04}).
Clearly, the orbit of the wide binary remains unconstrained.

\begin{table}
   \caption[]{Stellar properties of \object{HD 20782} and \object{HD 20781} }
     \label{t:hd20782}
      
       \begin{tabular}{lccc}
         \hline
         \noalign{\smallskip}
         Parameter   &  HD 20782 &  HD 20781  & Ref. \\
         \noalign{\smallskip}
         \hline
         \noalign{\smallskip}
$\mu_{\alpha}$ (mas/yr)  & 348.88$\pm$0.50    &  349.07$\pm$0.78     & 1 \\
$\mu_{\delta}$ (mas/yr)  & -64.82$\pm$0.73    & -67.80 $\pm$1.00     & 1 \\
RV     (km/s)            &   39.5$\pm$0.2     &  39.6 $\pm$ 0.2      & 2 \\ 
$\pi$  (mas)             &  27.76$\pm$0.88    &  27.86$\pm$1.23      & 1 \\
$R_{min}$ (kpc)          & \multicolumn{2}{c}{4.95}                  & 2 \\
$R_{max}$ (kpc)          & \multicolumn{2}{c}{8.16}                  & 2 \\
$ecc $                   & \multicolumn{2}{c}{0.24}                  & 2 \\
$z_{max}$ (kpc)          & \multicolumn{2}{c}{0.08}                  & 2 \\
V                        &  7.366             &  8.457               & 2 \\
ST                       & G3V                & K0V                  &  1  \\
${\rm Mass} (M_{\odot})$ &  0.90           &  0.84                   &  2 \\
                         &  1.00           &                         &  3 \\
                         & $0.969^{+0.024}_{-0.022}$     &             &  4 \\
Age$_{isoc}$  (Gyr)      &  13.0                       &             &  2  \\
                         &$7.1^{+1.9}_{-4.1}$ &                      &  3  \\
                         &$9.68\pm1.76$       &                      &  4  \\
Age$_{HK}$  (Gyr)        &   3-6            & 6                      &  5,6 \\

         \noalign{\smallskip}
         \hline
      \end{tabular}

References: 1 Hipparcos (ESA \cite{hipparcos});
            2 Nordstrom et al.~(\cite{nordstrom04});
            3 Valenti \& Fischer (\cite{vf05});  
            4 Takeda et al.~(\cite{takeda06});
            5 Jones et al.~(\cite{jones06});
            6 Gray et al.~(\cite{gray06})
\end{table}

We tested the dynamical stability of the system via numerical integration and
we have performed 10 simulation of this system following the evolution 
for $200$ Myr (577 binary orbits) using a RADAU integrator 
(Everhart et al.~\cite{1985dcto.proc..185E}).
The binary orbit was assumed with typical values of 
$a_{bin}=6000$~AU and $e_{bin}=0.60$,
the orbital parameter of the planet was chosen equal to the observed values, 
and inclination was chosen slightly inclined ($i = 5^\circ$) with respect 
to the binary orbital plane.
The planet results stable over  all 
the simulation, 
thus we argue that the large binary separation ensures dynamical 
stability of the planet 
in spite of its very high eccentricity.

Only two extrasolar planets have orbits with eccentricities larger than 0.8: 
\object{HD 20782b} and \object{HD 80606b}.
Both their host stars are member of wide common proper 
motion pairs.

While the link between the high eccentricity and binarity still needs 
statistical confirmation, nevertheless it is interesting to investigate
the possible ways in which a distant companion might have induced extreme
planet eccentricities.

As discussed in Sect.~\ref{s:discussion},
the extremely high eccentricity 
of  \object{HD20782b} is unlikely to be due to the
Kozai mechanism, as the period of the eccentricity oscillation
is much larger than the Hubble time.

An interesting possibility to explain both the very large separation
of the binary and the very high eccentricity of the planet orbiting
the primary is  a dynamical encounter of the binary
(or of an originally higher multiplicity system)
within a star cluster or with a passing star,
that might have perturbed both the binary and the planet orbit.
According to the simulations by Weinberg et al.~(\cite{weinberg87}), 
the probability of survival for a binary with an initial semimajor axis
of about 10000~AU after 4 Gyr is less than 50\%.
Furthermore, the actual semimajor axis
is probably larger than the observed projected separation by about
30\% (Duquennoy \& Mayor \cite{duq91}; Fischer \& Marcy \cite{fischer92}),
further decreasing the probability of survival, if this is the case
of \object{HD20782}.
The galactic orbit of \object{HD 20782} (Table \ref{t:hd20782}) 
makes the star  on average
closer to the Galactic center than the Sun, 
further increasing the chance of stellar encounters, because
of the higher stellar density.
Therefore, at the age of the system ($7.1^{+1.9}_{-4.1}$ Gyr;  
Valenti \& Fischer \cite{vf05}; $9.68\pm1.76$; Takeda et al.~\cite{takeda06}),
the probability that the currently observed orbit is not the original one
but it was significantly modified by dynamical encounters in the Galactic
disk is rather high.

The process of destruction of wide binaries
is typically the result of a large number of weak encounters
(see e.g. Portegies Zwart et al.~\cite{zwart97}).
The changes in the binary orbit caused by one or more stellar encounters 
might alter the orbit of the planet(s) orbiting one of the components.
A multi-planet system can be destabilized, leading
to chaotic evolution of the planetary orbits.
Dedicated dynamical modelling should be performed to check 
if the current configuration of the system might represent a possible
outcome of a stellar encounter. This is postponed to  future work.
The hypothesis of stellar encounters appears less appealing to explain
the extremely high eccentricity of \object{HD 80606b}, as in this case
the projected separation of the binary is about 1200~AU, 7-8 times
smaller than for \object{HD 20782}. However, the real orbit
remains unknown.

The possibility that the high planet eccentricity is unrelated to the
presence of the stellar companions and is caused by internal evolution of
the planetary system cannot be excluded.
Conclusive inferences on the origin of the very high eccentricity
of \object{HD 20782b} and \object{HD 86606b} are not possible from the available
data.
Further clues will come from follow-up observations, such as the search for 
further planetary and stellar companions, and from the discovery of further 
planets with very 
highly eccentric orbit and the study of the binarity of their hosts.


\subsection{HD 109749}
\label{s:hd109749}

\object{HD 109749} is listed in Hipparcos and CCDM catalogs as a binary 
(CCDM J12373-4049).
The Hipparcos magnitude of the secondary transformed to the standard system following
Bessell (\cite{bessell}) yields $V=10.76$, about 2.5 mag fainter than the primary
(V=8.09, Fischer et al.~\cite{fischer06}; see Table~\ref{t:hd109749}).
Table \ref{t:astrometry} shows the relative astrometry for the two
components. The constancy of the projected separation and position angle, 
coupled with the rather large proper motion of \object{HD 109749},
makes the physical association between the two objects very probable.
The projected separation of 8.35 arcsec corresponds to about 500 AU.

\begin{figure}
    \includegraphics[width=8cm]{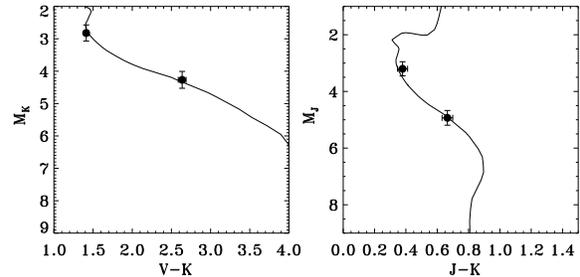}
      \caption{Absolute magnitudes and colors for the components of HD 109749
                (filled circles). Left panel: $M_K$ vs V-K; right panel:$M_J$ vs J-K.  
                 The 4 Gyr, Z=0.030 isochrone by Girardi 
                 et al.~(\cite{girardi02}) is overplotted. }        
         \label{f:hd109749_isoc}
   \end{figure}

To further check the physical association between the two stars, we identified 
them in 2MASS catalog (Cutri et al.~\cite{2mass}).
The photometry is shown in Table \ref{t:hd109749}.
When placed in a color-magnitude diagram, both components lies along the same isochrone,
indicating a common distance and  confirming the physical
association (Fig.~\ref{f:hd109749_isoc}).
The masses derived from the isochrone shown in Fig.~\ref{f:hd109749_isoc}
are 1.11 and 0.78~$M_{\odot}$.  

\begin{table}
   \caption[]{Stellar properties of \object{HD 109749} and \object{HD 109749B} }
     \label{t:hd109749}
      
       \begin{tabular}{lccc}
         \hline
         \noalign{\smallskip}
         Parameter   &  HD 109749 &  HD 109749B  & Ref. \\
         \noalign{\smallskip}
         \hline
         \noalign{\smallskip}

$\mu_{\alpha}$ (mas/yr)  &  -157.89$\pm$1.41   & -158                 & 1,2 \\
$\mu_{\delta}$ (mas/yr)  &    -5.48$\pm$1.25   & -6                   & 1,2 \\
$\pi$  (mas)             &   16.94$\pm$1.91    &                      & 1 \\
V                        &    8.09             & 10.76                & 4,5 \\
$J_{2MASS}$              &   7.057$\pm$0.021   & 8.788$\pm$0.024      & 6 \\
$H_{2MASS}$              &   6.797$\pm$0.031   & 8.289$\pm$0.051      & 6 \\
$K_{2MASS}$              &   6.678$\pm$0.024   & 8.123$\pm$0.024      & 6 \\
$M_{V}$                  &   4.23$\pm$0.25     &  6.90$\pm$0.25       & 1,5 \\ 
${\rm Mass} (M_{\odot})$ &   1.11              &  0.78                & 5 \\
${\rm Mass} (M_{\odot})$ &   1.2               &                      & 4 \\

         \noalign{\smallskip}
         \hline
      \end{tabular}

References: 1 Hipparcos (ESA \cite{hipparcos});
            2 USNO-B1.0 Catalog (Monet et al.~\cite{usno})
            3 Nordstrom et al.~(\cite{nordstrom04});
            4 Fischer et al.~(\cite{fischer06});  
            5 This paper;
            6 2MASS (Cutri et al.~\cite{2mass})

\end{table}
\begin{table}
   \caption[]{Relative astrometry of \object{HD 109749} }
     \label{t:astrometry}
      
       \begin{tabular}{lccl}
         \hline
         \noalign{\smallskip}
         Epoch  &  $\rho$ &  $\theta$  & Ref. \\
         \noalign{\smallskip}
         \hline
         \noalign{\smallskip}

  1881    &    8.2                &    179             & CCDM \\
  1987.35 &    8.35$\pm$0.04      &    179.76$\pm$0.21 & Sinachopoulos (\cite{sinach}) \\ 
  1991.25 &    8.353$\pm$0.014    &    180.0           & Hipparcos \\
  1999.27 &    8.28               &    179.75          & 2MASS \\

         \noalign{\smallskip}
         \hline
      \end{tabular}

\end{table}

\end{appendix}

\begin{appendix}

\setcounter{section}{1}

\section{White dwarf companions of planet hosts}
\label{s:wd}

Three planets are orbiting stars with white dwarfs
companions.
For binaries wide enough to escape common envelope
evolution, the mass loss from the originally
more massive star during the RGB and AGB phases determines
a widening of the orbit.
The mass loss phase lasts several Myr, longer than the typical 
binary orbital
periods.
In these conditions, the widening of the binary orbit
can be approximated by $a_{end}=a_{start}~M_{start}/M_{end}$, 
and no eccentricity
changes are expected (Lagrange et al.~\cite{lagrange06} 
and references therein).
In this section, we consider the three planet-host binaries
with a WD, placing constraints on the original
configuration of the system in terms of binary separation
and mass ratio.

\subsection{HD 13445 = GL 86}
\label{s:gl86}

The companion of \object{Gl86} was discovered by 
Els et al.~(\cite{els01}) and it was
classified as a brown dwarf on the basis of the near-infrared colors.
Mugrauer \& Neuhauser (\cite{mugneu05}) showed that the companion is instead 
a white dwarf.
Lagrange et al.~(\cite{lagrange06}) further constrained the properties
of the system, using archive radial velocity and new photometry and 
position measurements\footnote{The available observational constraints did not
allow Lagrange et al.~(\cite{lagrange06}) to derive a unique 
solution for the binary orbit. The orbit listed in Table~\ref{t:tablebin}
is a plausible, representative solution.}.
However, they also noted some inconsistency between the white dwarf
cooling age (1.2-1.8 Gyr depending on white dwarf mass and atmosphere
composition), the age of the star as derived from 
chromospheric activity (about 2-3 Gyr), and the presence of the planet. 
In fact, such a young age implies an original mass
of the companion that is too large to be compatible with
a dynamically stable orbit for the planet once the widening 
of the binary orbit
due to mass loss from the originally more massive star
is taken into account.

A picture that might explain the properties
of the system is the following.
Kinematic parameters ($U=-102$~km/s, $V=-75$~km/s, $W=-29$~km/s, 
galactic orbit 
$R_{min}=4.0$ kpc, $R_{max}=9.2$ kpc, $e=0.39$, $Z_{max}=0.4$ kpc; 
Nordstrom et al.~\cite{nordstrom04}) indicate a very 
old age\footnote{The difference between the kinematic and activity age
of \object{GL 86} was previously noted by 
Rocha Pinto et al.~(\cite{rochapinto02}).}.
The kinematic criterium of Bensby et al. (\cite{bensby05}) suggests
a probable membership to the thick disk.
Chemical abundances can also be used to constrain the galactic population.
At the metallicity of GL86 ([Fe/H]$\sim -0.20$),
differences between thin and thick stars are fairly small 
(the typical [$\alpha$/Fe] of thin disk and thick stars is 
about +0.05 and +0.15 respectively, Bensby et al. \cite{bensby05}).
High--resolution abundances analysis of several elements were published by
Allende Prieto et al.~(\cite{allende04}),  Valenti \& Fischer (\cite{vf05}),
Santos et al.~(\cite{santos04}), and
Gilli et al.~(\cite{gilli06}). 
The uncertainty in the chemical abundances of \object{GL 86} 
and the discrepancies between different studies do not allow
a firm classification, but they appear compatible 
with thick disk membership or with thick disk-thin disk transition
objects\footnote{The relative abundances
of Allende Prieto et al.~(\cite{allende04}) ([O/Fe]=+0.21, [Mg/Fe]=+0.13,
[Si/Fe]=+0.08, [Ca/Fe]=+0.18, [Ti/Fe]=+0.11, [Eu/Fe]=+0.30)
fit fairly well the expected abundance pattern for a thick star of 
the metallicity of {\object Gl86}, those by Gilli et al.~(\cite{gilli06})
and Ecuvillon et al.~(\cite{ecuvillon05})
are more ambiguous ([O/Fe]=-0.01, [Mg/Fe]=+0.21,
[Si/Fe]=-0.02, [Ca/Fe]=-0.10, [Ti/Fe]=+0.19). Valenti \& Fischer
(\cite{vf05}) find an  abundance ratio similar to that
of Allende Prieto et al.~(\cite{allende04}) for two elements
([Si/Fe]=+0.11, [Ti/Fe]=+0.10).}. 
Thick disk stars with this metallicity have typical ages of 
8 Gyr according to Bensby et al.~(\cite{bensby05}).
This is compatible with the lower limit (68\% confidence level) of 8.5 Gyr
estimated by Takeda et al.~(\cite{takeda06}) from isochrone fitting.
If \object{GL 86} is instead a thin disk star, 
its kinematic parameters strongly 
supports an old age.

Matching the cooling age of the white dwarf given by 
Lagrange et al.~(\cite{lagrange06})
with the assumed age of about 8 Gyr for the system
requires an original mass of about $1.2~M_{\odot}$.
Older ages yield lower initial masses.

A $1.2~M_{\odot}$ star ends its life as a $\sim 0.54~M_{\odot}$
white dwarf. The amount of mass loss implies a widening of the binary
orbit of about 40-50\% of its original separation.
To match the semimajor axis of 18.4~AU derived by 
Lagrange et al.~(\cite{lagrange06}),
the semimajor axis of the original orbit had to be about 13~AU.

A fraction of the mass loss by \object{GL86 B} during the
RGB and AGB evolution should have been
captured by the planet host.
A rough estimate based on the BSE binary evolution code by
Hurley et al.~(\cite{bse}), using the default
wind and accretion parameters, is 
$0.017~M_{\odot}$.
The $\sim0.017~M_{\odot}$ accreted mass corresponds to roughly half of 
the mass of the convective envelope of GL86
($M_{ce}=0.039\pm0.003$; Takeda et al.~\cite{takeda06}).
The lack of peculiar abundance of neutron capture
elements ([Ba/Fe]=--0.12; Allende Prieto et al.~\cite{allende04})
suggests an original  mass for the AGB star 
smaller than $1.3-1.5~M_{\odot}$ (Busso et al. \cite{busso99}),
in agreement with our estimate of $\sim 1.2~M_{\odot}$.

Along with mass, angular momentum should also 
have been transferred to the planet host.
This was  invoked to explain the high rotation rates
of companions of hot white dwarfs such as the barium dwarf 
\object{2RE J0357+283} (Jeffries \& Smalley \cite{jeffries96}).
The accretion of angular momentum by \object{GL86} would explain 
its activity level typical of a much younger 
star.

This scenario would be able to explain the available
observational constraints, including the stability of
the planetary orbit.
However, the formation of a planet with a companion at about 13~AU
(and with an eccentricity  of 0.4) represents a challenge
for current models (Hatzes \& Wuchterl \cite{hatzes05}).
The signature of accretion of angular momentum by \object{GL86}
is an indication that the binary system was in its current
status at the time of the mass loss by \object{GL86B}, about 1.5 Gyr ago.
Earlier modifications of the binary orbit due to e.g. dynamical
interactions within the native star cluster 
(Pfahl \& Muterspaugh \cite{pfahl06}) 
cannot be excluded.
The possible effects of the mass and angular momentum accretion 
on the mass and orbit of the planet remain to be investigated.

\subsection{HD 27442 = $\epsilon$ Ret}
\label{s:epsret}

\object{HD 27442}= $\epsilon$ Ret has a companion at a 
projected separation of about 13 arcsec = 240 AU.
According to Chauvin et al.~(\cite{chauvin06}), 
visual and  near-infrared photometry are not compatible
with any main sequence star. They then argue that the
companion is a white dwarf.
The  mass of the planet host yields a lower limit to
the original mass of the companion. 
Valenti \& Fischer (\cite{vf05}) determine a mass of 
$1.49~M_{\odot}$ (the value also adopted by B06), with
an age of 3.5 Gyr, while Takeda et al.~(\cite{takeda06})
list $M=1.48^{+0.22}_{-0.08}~M_{\odot}$ with an age of 
$2.84^{+0.60}_{-0.36}$ Gyr.
A lower limit to the WD mass
estimated using the BSE code (taking into account
the high metallicity of \object{$\epsilon$ Ret};
[Fe/H]=+0.42) 
is $M \sim 0.60~M_{\odot}$.
Theoretical models should be taken with caution at these
extremely high metallicities, as they fail to explain the
WD luminosity function of the old super-metal-rich open 
cluster \object{NGC 6791} (Bedin et al.~\cite{bedin05}). 
The relatively wide separation
from the primary (13 arcsec) and the moderately 
bright magnitude ($V \sim 12.5$, WDS) open the perspective for
a more detailed characterization of \object{$\epsilon$ Ret B}.
We can guess that the system  
had lost more than one third of its original mass,
implying a substantial widening of the binary orbit.
The original separation might have been about 150 AU, making the 
influence of the companion not negligible.

\subsection{HD 147513}
\label{s:hd147513}

\object{HD 147413} was classified as a member of the Ursa Major
moving group by Soderblom \& Mayor (\cite{soderblom83}) while King
et al.~(\cite{king03}) classified the membership as unprobable,
on the basis of the discrepant kinematic parameters. The
activity level and the position on the color-magnitude diagram
are instead fully compatible with membership.
Its companion (\object{WD 1620-39}=HIP 80300) is a widely studied
WD, for which Silvestri et al.~(\cite{silvestri01})
derived a mass of $0.65\pm0.01~M_{\odot}$ and a cooling age of 20 Myr.
The planet host was claimed to be a barium star ([Ba/Fe]=+0.37)
by Porto De Mello \& Da Silva (\cite{mello97}) but
Castro et al.~(\cite{castro99}) showed that the high barium
abundance is shared by other stars in the UMa moving group
(and possibly by other young stars). 
This points
toward a primordial abundance and not to accretion
by the WD companion, unexpected on the basis of the large
separation (more than 4000 AU).

We estimated the possible original mass of the (now) WD companion
using the BSE code (Hurley et al.~\cite{bse}).
Adopting the most recent age estimate of the UMa moving group 
(500 Myr; King et al.~\cite{king03})
and the cooling age by Silvestri et al.~(\cite{silvestri01}) leads to 
an original mass of $3.0~M_{\odot}$ and a WD mass of $0.75~M_{\odot}$, 
some $0.1~M_{\odot}$
larger than that derived by Silvestri et al.~(\cite{silvestri01}). 
The WD mass and cooling age by Silvestri et al.~(\cite{silvestri01})
can be matched simultaneously by adopting an initial mass of $2.3~M_{\odot}$
and a system age of 1 Gyr. This age is still marginally 
compatible with the 
photometry and activity level, if the star is not
a member of UMa moving group.
We conclude that the original separation of the binary was about a half
of the present one, as the system has lost about  half of its original mass.
Considering the current very wide separation, this does not change the
classification of this planet host as a very wide binary.

\end{appendix}


\begin{appendix}

\setcounter{section}{2}
\section{Individual objects}
\label{s:individuals}

We  further discuss  individual objects.

\begin{itemize}
\item
\object{HD 3651}: a brown dwarf companion was recently identified by
Mugrauer et al.~(\cite{mugrauer06b}). Its mass depends 
largely on the age of the system ($20-60~M_{J}$ for ages
of 1-10 Gyr respectively). Takeda et al.~(\cite{takeda06})
favor a very old age (older than 11 Gyr at a 68\% confidence
level). We then assume  $M=0.06~M_{\odot}$. Such a mass coupled
with the wide separation (480 AU) indicates that the 
classification of this object within the binary companions
and not within the 'super-planets' is appropriate.
The critical semiaxis for this system is uncertain as
Eq. 1 of Holman \& Wiegert (\cite{holman99}) is not defined
for such an extreme mass ratio.
\item
\object{HD 19994}: the visual orbit by Hale (\cite{hale94}) is very
preliminary.
\item
\object{HD 38529}: astrometric acceleration detected by Hipparcos.
According to Reffert \& Quirrenbach (\cite{reffert06}), 
this acceleration is due to the outer 
'planet' that has $m=37^{+36}_{-19}~M_{J}$.
The classification of this star as a two-planet host with a wide
stellar companion or a single-planet host with a brown dwarf and a further
wide companion is then ambiguous.
\item
\object{HD 41004}: the stellar metallicity of \object{HD 41004B} is not listed
by B06. We assume that of the primary.
\item
\object{HD 114762}: a companion with mass $0.075~M_{\odot}$ was found
by Patience et al.~(\cite{patience02}). Chauvin et al.~(\cite{chauvin06})
report a candidate companion, whose physical association has to be
established.
\item
\object{$\tau$ Boo}:
the very eccentric orbit by Hale (\cite{hale94}: $e=0.91$) 
implies a periastron of about 20 AU.
However, the latest position measurement by 
Patience et al.~(\cite{patience02}) indicates a larger
separation and similar position angle than
predicted by the preliminary orbit.
A redetermination of the binary orbit, taking into account
also the long term radial velocity trend reported by B06, would be
useful.
A further L dwarf companion candidate 
at  large separation has been reported by Pinfield et al.~(\cite{pinfield06}).
The physical association has to be confirmed.

\item
\object{HD 178911}: hierarchical triple system; a tight pair
(masses 1.10 and 0.79 $M_{\odot}$, semimajor axis 3 AU; 
Tokovinin et al.~\cite{tok00}) 
is at a projected separation of
about 640 AU from the planet host.
\item
\object{HD 186427}= 16 Cyg: hierarchical triple system; 16 Cyg A
(the companion of the planet host, at a projected separation of
850 AU) was shown to be a binary 
with a projected separation of about 70 AU (Patience et al.~\cite{patience02}).
The masses of 16 Cyg A and its companion are about 1.02 and 0.17 $M_{\odot}$
respectively. 
\item
\object{HD 189733}: this is the only host of a transiting planet
in a binary system.
The plane of the planet orbit is thus known.
According to Bakos et al.~(\cite{bakos06}),
from the available radial velocity and
position data of the binary system, there is evidence that
the binary and planetary orbit are not coplanar.
Therefore, this object is a candidate
for the occurrence of Kozai migration.
Characterization of the binary orbit will
allow to test this scenario.
\end{itemize}

\subsection{Objects not included in B06 catalog}
\label{s:notincluded}

The criteria for the inclusion of planets in the 
'Catalog of Nearby Exoplanets' are different  to 
other compilations such as the Extrasolar Planet Encyclopedia
maintained by J.~Schneider\footnote{www.exoplanet.eu}.
Three stars classified as planet hosts by the Extrasolar Planet 
Encyclopedia and not included in the Catalog of Nearby Exoplanets
are members of multiple systems:

\begin{itemize}
\item
\object{HD 188753A} was announced to host a hot Jupiter by Konacki
(\cite{konacki05b}). This star is a member of a hierarchical triple.
The close pair \object{HD 188753BC} reaches at periastron a
distance of just 6 AU from \object{HD 188753A}.
If confirmed, this candidate would be the planet in the closest
multiple system found up to now.
\item
\object{HD 196885}: a planet candidate was reported in 2004 at 
http://exoplanets.org/esp/hd196885/hd196885.shtml,
without successive publication of further details.
The star was found to be a tight binary (0.7 arcsec, 25 AU projected
separation) by Chauvin et al.~(\cite{chauvin06}).
\item
\object{HD 219449} is a giant star.
It was included in the study of R06 that showed that the system is
a hierarchical triple. A close pair, \object{HD 219430} (separation 18 AU) 
is at 2250 AU projected separation from the primary (the planet host
candidate).
\end{itemize}

\subsection{Additional unconfirmed binaries}
\label{s:unconfirmed}

A few additional planet hosts not included in Table \ref{t:tablebin}
have companion candidates whose identification or physical association 
needs confirmation:
These are: 
\begin{itemize}
\item
\object{HD 169830}:  a candidate companion at 11 arcsec was shown by 
R06 to have
photometric distance compatible with physical association with the planet host.
However, the low proper motion of the primary does not allow confirmation.
Further observations (radial velocity, spectral classification) are required.
\item
\object{HD 217107}: it is listed in WDS as a binary with a companion
with a projected separation of 0.3-0.5 arcsec=6-10 AU. The magnitude difference
is not provided.
According to R06, this detection needs confirmation.
Chauvin et al.~(\cite{chauvin06})  reported
null results of their adaptive optics search using NACO.
Their limit on $\Delta K_S$ at 0.3 arcsec is about 7.5 mag, corresponding 
to a mass of about $0.1~M_{\odot}$.
Vogt et al.~(\cite{vogt05}) also report  non detection of stellar companions
at separations larger than 0.1 arcsec.
The Hipparcos catalog does not include evidence for astrometric motion, 
expected for a stellar companion at 6 AU.
As noted by R06, the presence of a stellar companion at 6 AU would not be 
dynamically compatible with that of the outer planet ($a \sim 4$ AU). 
One possible explanation is that the system is seen nearly pole-on and 
that the stellar companion and the outer 'planet' 
are the same object. However, this possibility is not convincing considering 
that
the second epoch observation listed in WDS is 1997, fairly close to the 
periastron
of the RV orbit ($1998.7\pm0.3$). Detection should have been easier 
in the past few years, in contrast to the
null detections by deep adaptive-optics searches.
\item
\object{HD 168443}:
Reffert \& Quirrenbach (\cite{reffert06}) derived an astrometric mass 
of $36\pm12~M_{J}$ for the outer 'super-planet' \object{HD 168443c}.
As for \object{HD 38529}, there is then ambiguity about the classification
of the object.
\item
\object{HD 111232} and \object{HD 150706}: astrometric acceleration detected by Hipparcos (see R06).
\item
\object{HD 52265}, \object{HD 121504}, \object{HD 141937},  \object{HD 154857},
\object{HD 162020}, \object{HD 179949} and  \object{HD 183263}: first epoch
observations by Chauvin et al.~(\cite{chauvin06}) revealed companion
candidates. Follow-up observations are required to establish  
their physical association  with the planet hosts.
\end{itemize}

\end{appendix}

\begin{acknowledgements}

   This research has made use of the 
   SIMBAD database, operated at CDS, Strasbourg, France,
   and of data products from the Two Micron All Sky Survey. 
   We thank J.R.~Hurley for providing the BSE program.
   We thank G.~Takeda for useful information about the
   stellar ages of planet hosts.
   We thank R.~Gratton for his comments and suggestions.
   We thank the anonymous referee for a prompt and detailed
   report.
   This work was funded by COFIN 2004 
   ``From stars to planets: accretion, disk evolution and
   planet formation'' by the Ministero Univ. e Ricerca
   Scientifica Italy.

\end{acknowledgements}

\end{document}